\documentclass[lettersize,journal]{IEEEtran}
\usepackage{array}
\usepackage[caption=false,font=footnotesize,labelfont=rm,textfont=rm]{subfig}
\usepackage{textcomp}
\usepackage{stfloats}
\usepackage{verbatim}
\usepackage{graphicx}
\def\BibTeX{{\rm B\kern-.05em{\sc i\kern-.025em b}\kern-.08em
    T\kern-.1667em\lower.7ex\hbox{E}\kern-.125emX}}
\usepackage{balance}
\usepackage{color}
\usepackage{algorithmicx}
\usepackage{algorithm}
\usepackage[colorlinks,urlcolor=blue,linkcolor=blue,citecolor=blue]{hyperref}
\usepackage[T1]{fontenc}    
\usepackage{url}            
\usepackage{booktabs}       
\usepackage{amsfonts}       
\usepackage{nicefrac}       
\usepackage{microtype}      
\usepackage{xcolor}         
\usepackage{graphicx}
\usepackage{amsmath}
\usepackage{amsthm}
\usepackage{amssymb}
\usepackage{ifthen}
\usepackage{algpseudocode}
\usepackage{multirow}
\usepackage{ragged2e}

\begin{document}

\title{\textit{Silencer}: Robust Community Detection by Silencing of Noisy Pixels}

\author{Kai~Wu,~\IEEEmembership{Member,~IEEE,}
        Ziang~Xie,
        and~Jing~Liu,~\IEEEmembership{Senior~Member,~IEEE}
\thanks{This work was supported in part by the National Natural Science Foundation of China under Grant 62206205, in part by the Young Talent Fund of Association for Science and Technology in Shaanxi, China under Grant 20230129, in part by the Guangdong High-level Innovation Research Institution Project under Grant 2021B0909050008, and in part by the Guangzhou Key Research and Development Program under Grant 202206030003. (Corresponding author: Jing Liu.)}
}

\markboth{Journal of \LaTeX\ Class Files,~Vol.~18, No.~9, September~2020}%
{\textit{Silencer}: Robust Community Detection by Silencing of Noisy Pixels}

\maketitle

\begin{abstract}
Real-world networks carry all kinds of noise, resulting in numerous challenges for community detection. Further improving the performance and robustness of community detection has attracted significant attention. This paper considers edge noise, which causes edges in the network to be added or removed. Existing methods achieve graph denoising through link prediction or robustness in low eigenvectors. However, they are either limited in application scenarios or not determined for effectiveness. We find that the noisy pixel in the adjacency matrix has a certain proportion in the loss function, which makes the optimization of the community detection model seriously deviate from the correct direction. 
Thus, we design an flexible framework to silence the contribution of noisy pixels to loss function, called \textit{Silencer}. We take the nonnegative matrix factorization (NMF) and deep NMF methods as examples since they are the prime models for community detection. We first prove the convergence of \textit{Silencer} in NMF. Compared with existing methods, \textit{Silencer} show top performance in six real-world networks with random noise, adversarial perturbation, and mixed noise. Moreover, \textit{Silencer} works on random (ER), scale-free (BA), and small-world (WS) networks, and the improvement of \textit{Silencer} is gradually insignificant in the order ER, BA, and WS networks. 

\end{abstract}

\begin{IEEEkeywords}
Nonnegative matrix factorization, community detection, adversarial attack, network enhancement, self-paced learning.
\end{IEEEkeywords}

\section{Introduction}
\IEEEPARstart{C}{ommunity} detection (CD) is not only attractive for understanding the structures and functions of complex networks but also with increasing practical significance in many fields, such as social networks, coauthorship networks, and protein-protein interaction networks ~\cite{fortunato2010community,blondel2008fast}. For example, sponsors promote products on the Taobao shopping platform to target users in discoverable communities. A great deal of effort has been devoted to developing community detection methods, such as spectral clustering ~\cite{gennip:spectral}, nonnegative matrix factorization (NMF)~\cite{lee1999learning,guan2022community,chaobao:NMF-survey,Ma2017Evolutionary,Ma2019Community,guan2023community}, and deep learning for community detection ~\cite{liu:survey,Jin2021Survey,su2022comprehensive}. However, their capability to detect the true communities confronts numerous challenges because they strongly depend on the topological structure of the underlying network, which is vulnerable in real-world scenarios.

There are two streams of work to disturb the original network. The objective of intentional obfuscation and privacy-preserving work is to prevent the adversary from achieving successful network analysis.
The goal of concealing individuals and communities in a social network is common \cite{waniek2018hiding}. It helps the public safeguard their privacy from government and corporate interests. 
In particular, the goal of adversarial attacks against community detection is to conceal target communities, or sensitive edges \cite{waniek2018hiding,li2020adversarial,fionda2017community,chen2019ga}. It finally generates specific noisy (adversarial) networks, which can significantly affect the performance of community detection methods, leading to community detection deception. The other is missing information or inaccurate measurements, leading to noise in the network \cite{he:self}. Against this background, we focus on promoting the performance of community detection against noise.

These robust community detection methods can be divided into two parts. The first one employs the link prediction methods to predict missing edges, and then the repaired network is employed for community detection methods \cite{tran2021community,shao2019community}. However, this strategy is challenging to handle noisy networks with added edges. The second is the network enhancement methods, which improve community detection methods by weighting or rewiring networks \cite{kang2022community,li2020community}. These methods do not focus on the noisy network but only strengthen the network for community detection. Most of them focus on modularity-based or motif-aware community detection. These enhancement methods do not develop the corresponding strategy to cope with the noisy network, resulting in the inability to handle noise. Therefore, designing a robust community detection scheme is significantly urgent.

This paper takes another idea to enhance community detection algorithms against noisy networks, considering the hybrid scene of adding and removing edges and fusing the characteristics of community detection algorithms. We first define the adding and removing edges as \textbf{noisy pixels} (the elements in adjacency matrix). The existence of noisy pixels causes the performance of the community detection algorithm to deteriorate. Thus, an intuitive solution is to find these noisy pixels and modify them. A more general solution is reducing the weight of the role played by noisy pixels in community detection methods, that is, reducing their contributions to the total loss function. However, in many real-world applications, identifying clean and noisy samples in a given training dataset may be onerous or conceptually tricky.

Our idea is inspired by the phenomenon that noisy pixels cause the loss value of a model to become more significant. The pixel-level loss corresponding to the noisy pixel increases, and the probability is greater than that of the clean pixel. 
The following questions are 1) how to set the threshold to classify noisy and clean pixels and 2) how to calculate the pixel-level loss. Fortunately, we borrow the age parameter employed in self-paced regularization~\cite{zhao:self,kumar:paced,jiang:diversity,meng:theoretical} to address the first question by treating clean edges as easy samples and noisy edges as hard samples. \textit{Silencer} trains the community detection model on easy samples first and then gradually adds complex samples into consideration. Any community detection that can calculate the pixel-level loss can be employed. This paper also takes the NMF \cite{chaobao:NMF-survey} and deep autoencoder-like NMF (DANMF)~\cite{ye:autoencoder} as the example to calculate the pixel-level loss. Although NMF-based community detection methods have also been widely adopted due to their versatility, flexibility, simplicity, effectiveness, and interpretability \cite{chaobao:NMF-survey}, the performance of them sharply dropped when dealing with noisy networks.

As shown in the experiments conducted on six real-world networks with random noise, adversarial attack, and mixed noise, NMF and DANMF have poor performance, but joining \textit{Silencer} and the original has a strong resistance to noisy networks. Moreover, the systematic comparison with the existing community detection methods handling noisy networks demonstrates the excellent performance of \textit{Silencer}. We also show the learned weights of \textit{Silencer} and find that reducing the contribution of noisy pixels to the total loss can effectively improve the anti-noise ability of community detection methods.

The highlights of this paper are summarized as follows:
\begin{itemize}
    \item We present a general robust community detection framework to address a variety of edge noises, such as random noise, adversarial attack, and mixed noise. \textit{Silencer} can keep a stable performance gain even in loud and complicated noise environments, which is challenging for current methods. \textit{Silencer} is suitable for a range of methods that can calculate the pixel-level loss.
    \item We take the NMF and DANMF as examples. We first prove the convergence of \textit{Silencer} in NMF. Moreover, \textit{Silencer} outperforms the state-of-the-art methods in six networks with different noises. \textit{Silencer} achieve a high gain on random, scale-free, and small-world networks, which shows our proposal has a very wide range of application scenarios.
\end{itemize}

The rest of this paper is organized as follows. Section II summarizes the related work on DNMF-based community detection methods, robust community detection methods, and network enhancement methods. Section III formalizes the studied problem and introduces the background of DNMF. Section IV presents the \textit{Silencer} approach to the NMF and DANMF models. Section V conducts extensive experiments to evaluate the effectiveness of \textit{Silencer}. 
Finally, Section VI concludes the entire paper.

\section{Related Work}

This section introduces the existing NMF-based community detection methods and the effort of community discovery from noisy networks to demonstrate the motivation of \textit{Silencer}.

\subsection{DNMF-based Community Detection}

Community detection has always been one of the most popular research topics. Its purpose is to identify communities in complex graph structure data. Although there is no unified definition of communities, a consensus can be reached that points within communities are more closely connected than those between communities ~\cite{fortunato2010community}. A large class of heuristic algorithms with measurement such as metric-based heuristic algorithms~\cite{newman2006modularity,shi2000normalized}, and statistical inference-based approaches~\cite{nowicki2001estimation}, spectral clustering~\cite{gennip:spectral}, and NMF ~\cite{liu:survey,wang2017community} . In recent years, there have been studies on community detection by the graph embedding method, which has also achieved good results~\cite{cavallari2017learning,grover2016node2vec,tang2015line,wang2017community}.

NMF-based community detection approaches have also been widely adopted due to their advantages in versatility, flexibility, simplicity, and interpretability~\cite{liu:survey,yang2013overlapping}. NMF is used to handle the community detection problem based on the assumption that links are far more likely to occur between nodes in the same cluster. The adjacency matrix of complex networks is decomposed into two matrices, the mapping $\mathbf{U}$ and the community membership matrix $\mathbf{V}$. 
Moreover, the NMF method has robust scalability for overlapping community detection by setting the threshold value of participation and introducing corresponding prior knowledge by designing a feature matrix and regular terms.
Due to the shallow structure of NMF-based community detection approaches, it is still a considerable challenge to detect communities from complex networks~\cite{liu:survey}. To overcome this issue, Ye \emph{et al.}~\cite{ye:autoencoder} proposed DANMF to discover communities that contains multilayer mapping between the cluster membership space and the network space. Consequently, two DANMF-based community detection methods~\cite{esraa2023community,he2020network} attempts to cope with noisy networks and enhance the robustness of DANMF by using the $l_{2,1}$ norm. However, the performance of DANMF dropped when handling noisy networks. To overcome the problem of DANMF in handling noisy networks, the proposed \textit{Silencer} takes edges gradually from easy to hard into consideration during training rather than considering all pixels equivalently.

\subsection{Robust Community Detection}
\subsubsection{Community Detection with Link Prediction}

Generally speaking, there is a basic assumption that the input network is clean, but this assumption must be validated. The studies ~\cite{gabielkov2012complete,huisman2009imputation,borgatti2006robustness,he:self} have shown that the noise of the input matrix affects the critical indicators of the network, which also increase the difficulty of community detection.

Several studies on the network edge-missing problem have emerged in recent years by combining the link prediction problem with the community discovery problem. Tran \emph{et al.}~\cite{tran2021community} employed the link prediction method to recover both missing nodes and edges before performing overlapping community detection on the incomplete network. Moreover, Shao \emph{et al.}~\cite{shao2019community} developed a unified framework to conduct link prediction and community detection simultaneously. However, these methods assume that the more missing edges the link prediction methods can predict correctly, the more accurate the community detection is. However, the assumption may not hold in reality as not all edges play the same role in community detection. To bridge the gap between community detection and link prediction, He \emph{et al.}~\cite{he:self} designed a community self-guided generative model which jointly fills the edges-missing network and detects communities.
Zhu \emph{et al.}~\cite{zhu2023unified} employs adversarial training framework to enhance the robustness of many existing NMF-based community detection methods and this framework is suitable to edge-missing networks. These studies only analyzed the edges by deleting them. However, the noise in a network is from the loss of edges and the misconnected edges. We propose a more practically feasible approach to cope with this issue.

\subsubsection{Network Enhancement for Community Detection}

Existing network enhancement approaches enhances existing community detection algorithms by preprocessing networks via weighting or rewiring. 

They employed the random walk-based strategies \cite{PhysRevELai,DEMEO2013648} or a series of edge centrality indices \cite{SUN2014346} to update the edge weights. For example, Meo \emph{et al.} \cite{DEMEO2013648} designed a weighting algorithm by introducing a k-path edge centrality measure to compute the centrality as edge weight. Since these methods cannot replace missing edges in an incomplete network, they may fall short when there are missing edges.
Li \emph{et al.} \cite{EdMot2019} developed an edge enhancement method, which employed motif information in networks for motif-aware community detection. 
He \emph{et al.} \cite{HE2016602} tended to integrate the network enhancement into the entire NMF procedure to detect communities. It can utilize the topological similarity more efficiently because it does not use network enhancement as a preprocessing step before community detection. Zhou \emph{et al.} \cite{RobustECD} proposed two strategies to enhance network for a variety of community detection methods: 1) the first one was inspired by community structures, which shows a dense connection in intra-communities and a sparse one in inter-communities; 2) the second one employed link prediction to enhance network structure by complementing missing edges or predicting future edges. Kang \emph{et al.} \cite{kang2022community,kang2020cr} proposed a preprocessing stage for strengthening the community structure of the clean network. This function can be realized by adding non-existent predicted intra-community edges and deleting existing predicted inter-community edges. Moreover, to address the problem of community detection in adversarial multiview networks, Huang \emph{et al.} \cite{huang2021higher} employed the higher order connection structure to enhance the intra-community connection of each view. NetRL \cite{xu2021netrl} turns the problem of network denoising into edge sequences generation, which can be solved by deep reinforcement learning. NetRL takes advantage of the node classification task to guide this process. However, this task-specific objective is difficult to obtain.

Most of them perform poorly in strong noise cases. At the same time, the above network enhancement methods aim at methods such as modularity or motif-aware community detection. Unlike the above techniques, we try to solve the problem of robust community detection by silencing noisy pixels. The proposed methods are effective in handling strong noise with different network sizes.

\subsection{Task-unaware Network Enhancement}

Some network enhancement (denoising) efforts are unrelated to the community detection task. Feizi \emph{et al.} \cite{feizi2013network} considered the network denoising problem as the inverse of network convolution. Then, they exploited eigendecomposition to remove the combined effect of all indirect paths. Wang et al. \cite{wang2018network} improved the signal-to-noise ratio of undirected, weighted networks based on diffusion algorithms. They think nodes connected by paths with high-weight edges are more likely to have direct and high-weight edges. Mask-GVAE \cite{li2021mask} used the robustness in low eigenvectors of graph Laplacian against random noise and decomposed the input network into several stable clusters. Xu \emph{et al.} \cite{xu2022robust} exploited subgraph characteristics and graph neural networks to perform missing link prediction/detection, which can enhance the quality of the network. However, this method needs the supervision of clean networks.

\section{Preliminaries}


\subsection{Problem Definition}

In this paper, the matrix will be shown in bold capital letters. For matrix $\mathbf{M}$,
${\lbrack\mathbf{M}\rbrack}_{bd}$, ${\parallel \mathbf{M} \parallel}_F$, $\mathbf{M}^T$, and
$tr(\mathbf{M})$ denote the $d$th element of the $b$th row of the $ \mathbf{M}$ matrix, the Frobenius norm of $\mathbf{M}$, the transpose of $ \mathbf{M}$, and the trace of $\mathbf{M}$, respectively. The algorithm in this paper is a decomposition of the non-negative feature matrix. 

Formally, consider an undirected network $\mathbf{G = (Ve, A, E)}$, where $\mathbf{Ve} = \{ve_1, \cdots, ve_n\}$ denotes the set of nodes in the network (with $|\mathbf{Ve}|=n$) and $\mathbf{E}=\{e_{ij}|ve_i, ve_j \in \mathbf{Ve}\}$ indicates the set of edges. Without specification, the feature matrix is the adjacency matrix $\mathbf{A} \in \mathbb{R}_{+}^{n \times n}$ of the input network, and $n$ is the number of nodes. The subsequent experiments used are undirected networks without attribute and weight, i.e., ${\lbrack\mathbf{A}\rbrack}_{bd}$ has only two values of 0 and 1. In this paper, we focus on studying noisy networks, where some added edges that should not exist in $\mathbf{E}$, while a few meaningful and correct edges are removed in $\mathbf{E}$. 

\textbf{Definition 1 Noisy Pixel.} \textit{Given a noisy network $\mathbf{G=(Ve, A, E)}$, noisy pixel ${\lbrack\mathbf{A}\rbrack}_{bd}$ is the position of noise, where the added or removed edge appears.}

\noindent The goal of this paper is to study how to enhance $\mathbf{G}$ for accurate community detection, which is formulated as follows:

\textbf{Definition 2 Robust Non-overlapping Community Detection.} \textit{Given a noisy network $\mathbf{G=(Ve, A, E)}$, the goal of robust Non-overlapping community detection is to decrease the negative effect of noisy pixels on non-overlapping community detection so as to obtain accurate communities.}

\subsection{NMF and DANMF models}



Suppose that two nonnegative matrices $\mathbf{U} \in \mathbb{R}^{n \times k}$ and $\mathbf{V} \in \mathbb{R}^{k \times n}$, where each column of $\mathbf{U}$ denotes the description of a community, each column of $\mathbf{V}$ represents the association relationship of a node to different communities, and $k$ is the number of communities. Then, ${\lbrack\mathbf{U}\rbrack}_{bl}{\lbrack\mathbf{V}\rbrack}_{ld}$ can be interpreted as the contribution of the $l$th community to the edge ${\lbrack\mathbf{A}\rbrack}_{bd}$. NMF is to make ${\lbrack\mathbf{UV}\rbrack}_{bd}$ as close to ${\lbrack\mathbf{A}\rbrack}_{bd}$ as possible, which minimizes the following objective function:
\begin{equation}\label{eq:1}
\min_{\mathbf{U},\mathbf{V}}\mathcal{L} = {\parallel \mathbf{A} - \mathbf{UV} \parallel}_{F}^{2}, \quad
\mathrm{\text{s.t.,}}  \quad \mathbf{U} \geq 0, \mathbf{V} \geq 0.
\end{equation}
Unlike the NMF-based community detection approach, DANMF \cite{ye:autoencoder} is based on deep self-encoders, which considers the adjacency matrix decomposition as an encoding process and employs the idea of self-encoder symmetry. $\mathbf{U}$ should also be able to map $\mathbf{A}$ to the community membership space $\mathbf{V}$, i.e.$\mathbf{V} =\mathbf{U}^{T}\mathbf{A}$. Inspired by the idea of a deep network, $\mathbf{U}$ is further decomposed, thus obtaining the mapping between the network and the community member space containing as much complex hierarchical and structured information as possible, which results in the objective function of the DANMF algorithm:
\begin{equation}\label{eq:5}
\begin{split}
\min _{\mathbf{U}_{i}, \mathbf{V}_{p}} \mathcal{L} & =\mathcal{L}_{\mathrm{D}}+\mathcal{L}_{\mathrm{e}}+\lambda \mathcal{L}_{\mathrm{reg}} \\
&= {\|\mathbf{A} - \mathbf{U}_1\mathbf{U}_2\cdots\mathbf{U}_p\mathbf{V}_{p}\|}_F^2 \\
&+ {\|\mathbf{V}_{p} - \mathbf{U}_{p}^{T}\cdots\mathbf{U}_{2}^{T}\mathbf{U}_{1}^{T}\mathbf{A}\|}_{F}^{2} +  \lambda\operatorname{tr}\left( \mathbf{V}_{p}\mathbf{{LV}}_{p}^{T} \right)
\end{split}
\end{equation}

\[\begin{matrix}
\mathrm{\text{s.t.\ }}  \mathbf{V}_{p} \geq \mathbf{0},\mathbf{U}_{i} \geq \mathbf{0},\forall i = 1,2,\cdots,p.
\end{matrix}\]
where $\mathcal{L}_{reg}$ is the regularization term and further respect the intrinsic geometric structure of node pairs. $\lambda > 0$ is the regularization hyper-parameter. Unless otherwise specified, $\mathbf{V}_{p} \in \mathbb{R}^{k \times n} \geq \mathbf{0}$ represents the matrix ${\lbrack\mathbf{V}_{p}\rbrack}_{bd} \geq 0$, and $\mathbf{L}$ is the Laplacian matrix. We set $\mathbf{L = D - A}$, where $\mathbf{D}$ is a diagonal matrix whose diagonal elements are the corresponding row sums of $\mathbf{A}$. The adjacency matrix $\mathbf{V}_{p}$ is factorized into $p+1$ nonnegative factor matrices $\mathbf{A} \approx \mathbf{U}_1\mathbf{U}_2\cdots\mathbf{U}_p\mathbf{V}_{p}$, where $\mathbf{U}_i \in \mathbb{R}^{r_{i-1} \times r_i} (1 \leq i \leq p)$ and $n=r_0 \geq r_1 \geq \cdots r_p = k$. By Optimizing $\mathcal{L}_{\mathrm{D}}$, we obtain more accurate community detection results $\mathbf{V}_{p}$. Moreover, similar to the encoder component in autoencoder, $\mathcal{L}_{\mathrm{E}}$ is to project $\mathbf{A}$ into the community membership space $\mathbf{V}_{p}$ with the aid of the mapping $\mathbf{U}_i$.

\begin{figure}[ht]
\centering
\includegraphics[width=1\linewidth]{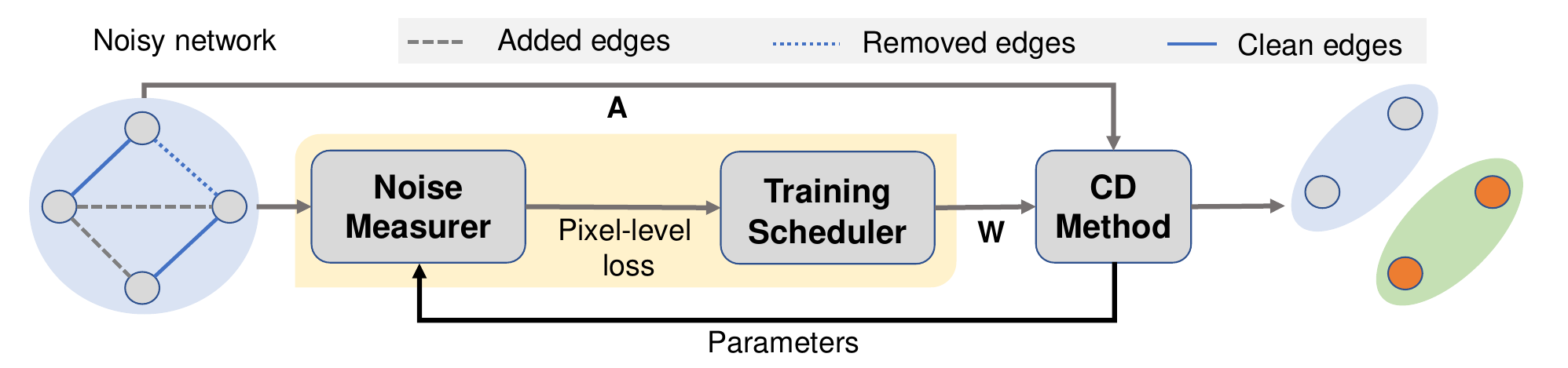}
\caption{Schematic depiction of \textit{Silencer}. The Noise Measurer aims to calculate the pixel-level loss based on the feedback of the CD method. Training Scheduler updates the weights for each pixel. The part of the CD method can be any community detection method that can obtain pixel-level loss.}
\label{fig:2}
\end{figure}

\section{\textit{Silencer}}

This section first describes the base framework of \textit{Silencer} and then shows the detailed application to NMF and DANMF models. Here, we first define the Pixel-level Loss used for the following parts.

\textbf{Definition 3 Pixel-level Loss.} \textit{Given the loss $\mathcal{L}$ after the given algorithm stops. Pixel-level loss $\mathcal{L}$ is the sum of the losses contributed by each pixel ${\lbrack\mathbf{A}\rbrack}_{bd}$, where $b,d = 1,2,\cdots,n$.}

The main process of \emph{Silencer} is demonstrated as follows:
\begin{itemize}
    \item \textbf{Step 1}: The employed base model outputs communities using the weight matrix $\mathbf{W}$ and the original network.
    \item \textbf{Step 2}: The noise measurer reconstructs the network using the community information to get the pixel-level loss.
    \item  \textbf{Step 3}: The training scheduler updates the weight matrix W using the pixel-level loss.
    \item \textbf{Step 4}: Repeat steps 1 through 3 until the model converges and outputs the final communities.
\end{itemize}

\subsection{\textit{Silencer} in NMF}
This section takes NMF as an example. \textit{Silencer} is to reduce the contribution of pixels with significant loss to the total loss (\ref{eq:1}), i.e., silencing noisy pixels.
\subsubsection{Model}
To achieve this goal, it is the most straightforward idea to embed the weight $\mathbf{W}$ into NMF by treating $\mathbf{W}$ as a variable describing the noise strength of each edge. If ${\lbrack\mathbf{W}\rbrack}_{bd}=0$, the difference between ${\lbrack\mathbf{UV}\rbrack}_{bd}$ and ${\lbrack\mathbf{A}\rbrack}_{bd}$, called pixel-level loss, is strictly set to 0. The process of calculating pixel-level loss can be considered as \textbf{Noise Measurer}.

Then, we can train NMF and select pixels. With the NMF model obtained from the selected pixels, we can compute the pixel-level loss for each edge in the entire network. The more significant the pixel-level loss, the more likely it is to be a noisy pixel. Then, the purpose of silencing noisy pixels is achieved by reducing the influence of the pixels with the larger loss on the NMF loss function. If one pixel ${\lbrack\mathbf{A}\rbrack}_{bd}$ is identified as noise, ${\lbrack\mathbf{W}\rbrack}_{bd}$ should be set to 0; otherwise 1. 
The process stops until \textit{Silencer} converges and all edges are considered. We employ self-paced regularization to act as \textbf{Training Scheduler} by gradually reducing its age parameter and 
 updating weights. This is based on the assumption that optimizing noisy pixels is more challenging than cleaning pixels.

The loss function of \textit{Silencer} in NMF is shown as follows:
\begin{equation}\label{eq:sp1}
\begin{gathered}
\min\limits_{\mathbf{U},\mathbf{V},\mathbf{W}}\mathcal{L}_{nmf}  =   \sum_{b,d = 1}^{n,n}{\left\{ \left [ \mathbf{W} \right ]_{bd}\left( \left\lbrack \mathbf{A} \right\rbrack_{bd} - \left\lbrack  \mathbf{U} \mathbf{V} \right\rbrack_{bd} \right)^{2} \right\}}\\ + f(\mathbf{W},\gamma)+ \lambda\operatorname{tr}\left( \mathbf{V}\mathbf{LV}^{T} \right) 
 \\
 \mathrm{\text{s.t.  }}  \mathbf{V},\mathbf{U} \geq \mathbf{0},  \mathbf{W} \in {\lbrack0,1\rbrack}^{n \times n},
 \end{gathered}
\end{equation}
where ${\lbrack\mathbf{W}\rbrack}_{bd}$ represents the weights for the contribution of each edge to the loss $\mathcal{L}$. For $f(\mathbf{W},\gamma)$, we give the soft weight operator to use for testing in this paper.
\begin{equation}\label{eq:9}
 f(\mathbf{W},\gamma) = \sum_{b,d}^{n,n}{}\frac{1}{{\lbrack\mathbf{W}\rbrack}_{bd}+1\mathbf{/}\gamma}.
\end{equation}

\subsubsection{Optimization}

If the pixel-level loss $l = \left( \left\lbrack \mathbf{A} \right\rbrack_{bd} - \left\lbrack  \mathbf{U} \mathbf{V} \right\rbrack_{bd} \right)^{2}$ is greater than the threshold, it is determined as a noisy pixel, and ${\lbrack\mathbf{W}\rbrack}_{bd}=0$; otherwise, it is the clean pixel and ${\lbrack\mathbf{W}\rbrack}_{bd}=1$. 
Therefore, we need to reduce the contribution of this pixel-level loss to the loss $\mathcal{L}$. We use self-paced learning regularization to determine the threshold.

The alternative optimization strategy ~\cite{kumar:paced} is used to solve the problem. We first solve the following problem with fixed $\mathbf{U,V}$: 
\[\mathbf{W} = \arg\min_{\mathbf{W}}\mathcal{L}_{nmf}\]
The update formula (\ref{eq:13}) for $\mathbf{W}$ are obtained from the soft weight operator (\ref{eq:9}) and the loss function (\ref{eq:sp1}).
\begin{equation}\label{eq:13}
{\lbrack\mathbf{W}\rbrack}_{bd} = \left\{
\begin{matrix}
1 & l_{bd} \leq (\frac{\gamma}{\gamma + 1}) \\
0 & l_{bd} \geq \gamma^{2} \\
\left(\frac{1}{\sqrt{l_{bd}}} - \frac{1}{\gamma}\right) & \mathrm{\text{\ otherwise\ }} \\
\end{matrix} \right.
\end{equation}
 where \(l_{bd} = {(\left\lbrack \mathbf{A} \right\rbrack_{bd} - {\left\lbrack \mathbf{UV} \right\rbrack_{bd}})}^{2}\).
The second problem is shown as follows:
\[\mathbf{U,V} = \arg\min_{\mathbf{U,V}}\mathcal{L}_{nmf}\]
We use (\ref{eq:sp2}) and (\ref{eq:update V}) to update $\mathbf{U}$ and $\mathbf{V}$.
\begin{equation}\label{eq:sp2}
\mathbf{U} \gets \mathbf{U} \odot 
\frac
{\left ( \mathbf{W} \odot \mathbf{A}  \right )\mathbf{V}^{T}  }
{\left ( \mathbf{W} \odot\left [ \mathbf{UV}  \right ]  \right ) \mathbf{V}^{T} } 
\end{equation}
\begin{equation}\label{eq:update V}
\mathbf{V} \gets \mathbf{V} \odot 
\frac
{\left (\mathbf{U}^{T} \left [ \mathbf{W} \odot \mathbf{A}  \right ]  \right ) +\lambda \mathbf{VA}   }
{\left  (\mathbf{U}^{T} ( \mathbf{W} \odot\left [ \mathbf{UV} \right ]  \right ) +\lambda \mathbf{VD} }
\end{equation}
where $\odot$ denotes the element-wise product.

\subsubsection{Convergence Analysis}

We prove its convergence by exploring the relationship between the update strategy and Majorization Minimization (MM) scheme. For the convenience of expression in the subsequent exposition, we denote the variables \(\mathbf{U, V}\) as \(\theta\), and $l=\sum_{b,d}{l_{bd}} $ where $l_{bd}$ is the pixel-level loss. We construct \(F(l) = \int_{0}^{l}{}\mathbf{W}^{*}(l,\gamma)dl\), where $\mathbf{W}^{*}(l,\gamma) = \arg\min_{\mathbf{W}}\sum_{b,d}{\mathbf{W}_{bd}l_{bd} + f(\mathbf{W}_{bd},\gamma)}$. Let \(l_{\text{bd}} = {(\mathbf{A}_{\text{bd}} - {\lbrack\mathbf{\text{UV}}\rbrack}_{\text{bd}})}^{2}\), we can find \(\mathbf{W}^{*}(l, \gamma) = \arg\min_{\mathbf{W}}L_{\text{Snmf}}\). And then, using the Taylor series and the concave function property, we can get the following upper bound $Q^{\left( bd \right)}\left( \theta \middle| \theta^{*} \right)$
of \(F\left( l_{bd}\left( \theta \right) \right)\):
\begin{equation*}
\begin{split}
Q^{\left( bd \right)}\left( \theta \middle| \theta^{*} \right) &= F_{}\left( l_{bd}\left( \theta^{*} \right) \right) + \mathbf{W}_{bd}^{*}\left( l_{bd}\left( \theta \right) - l_{bd}\left( \theta^{*} \right) \right) \\
&\geq F_{}\left( l_{bd}\left( \theta \right) \right)
\end{split}
\end{equation*}
Calculating \(\mathbf{W}^{*}(l,\gamma)\) to obtain
\(Q(\theta|\theta^{*}) = \sum_{b,d}^{}{Q^{(bd)}(\theta|\theta^{*})}\)
and minimizing \(Q(\theta|\theta^{*})\) to minimize
\(\sum_{b,d}^{}{F\left( l_{\text{bd}}\left( \theta \right) \right)}\)
are the standard MM algorithm framework. This is consistent with our update strategy, which can ensure its convergence. The details process are shown as shown follows:
\begin{equation*}
\begin{split}
\theta = arg\min_{\theta}Q\left( \theta \middle| \theta^{*} \right) &= arg\min_{\theta}\sum_{b,d}^{}\left( \mathbf{W}_{bd}^{*}l_{bd}\left( \theta \right) \right)\\
&= \arg\min_{\theta}\mathcal{L}_{nmf}
\end{split}
\end{equation*}

\subsection{\textit{Silencer} in DANMF}
We also take the DANMF model as an example to demonstrate the effectiveness of \textit{Silencer}. As shown in \cite{ye:autoencoder}, DANMF has significant performance improvement compared with the shallow structure of the NMF-based model. Thus, we focus on the application of \textit{Silencer} to DANMF. We compare the performance of DANMF on a network with noise. We observe that as the intensities of noise increase, there is a significant decrease in the performance of DANMF by the encoder and even drags down the DANMF performance. To apply \textit{Silencer} to DANMF, we need to solve the following two critical questions:
\begin{itemize}
    \item \textbf{Q1: Noise Measurer}: how to design and calculate the pixel-level loss;
    \item \textbf{Q2: Training Scheduler}: how to update the weights for the DANMF model.
\end{itemize}
The process of answering \textbf{Q1} and \textbf{Q2} is to explain \textbf{Noise Measurer} and \textbf{Training Scheduler}. See \textbf{AQ1} and \textbf{AQ2}.
\subsubsection{Model}

The straightforward solution is to set weights on both $\mathcal{L}_{\mathrm{D}}$ and $\mathcal{L}_{\mathrm{e}}$ to silence noisy edges. However, we partially embed the encoder into the \textit{Silencer} due to two considerations. 

1) Because of the deep network, the noise is accumulated and amplified layer by layer as it is passed in. This case causes the use of the encoder to reconstruct a $\mathbf{V}_{p}$ with a significant error, making DANMF have more noise being passed in compared to DANMF without an encoder. 

2) The most significant feature of network data, sparsity, causes $\mathbf{W} \odot\ \mathbf{A}$ to accumulate zero elements during the update process continuously, which eventually drives $\mathbf{U}$ to a meaningless all-zero matrix. Thus, similar to it, setting weights to $\mathcal{L}_{\mathrm{D}}$ fails.

Our strategy allows \textit{Silencer} to retain the advantages of the encoder for community discovery in DANMF while keeping robustness to noisy inputs.\\ \textbf{AQ1}: we design $\mathcal{L}_E^{'}$ to answer \textbf{Q1}. Based on the above reasoning, we introduce (\ref{eq:8}) from $\mathcal{L}_{E}$.

\begin{equation}\label{eq:8}
\begin{gathered}
\min _{\mathbf{U}_{i}, \mathbf{V}_p, \mathbf{W}} \mathcal{L}  =\mathcal{L}_{\mathrm{D}}+\mathcal{L}_E^{'}+\lambda \mathcal{L}_{\mathrm{reg}} + f(\mathbf{W},\gamma) \\
\mathcal{L}_E^{'} = \sum_{b,d = 1}^{k,n}{\left\{ \left [ \mathbf{W} \right ]_{bd}\left( \left\lbrack \mathbf{V}_{p} \right\rbrack_{bd} - \left\lbrack \mathbf{U}_{p}^{T}\ldots\mathbf{U}_{1}^{T}\mathbf{A} \right\rbrack_{bd} \right)^{2} \right\}} \\
 \mathrm{\text{s.t.  }}  \mathbf{V}_p,\mathbf{U}_i \geq \mathbf{0}, \forall i = 1,\cdots,p, \mathbf{W} \in {\lbrack0,1\rbrack}^{k \times n},
 \end{gathered}
\end{equation}
where ${\lbrack\mathbf{W}\rbrack}_{bd}$ represents the weights for the contribution of each edge to the loss $\mathcal{L}_E$. The pixel loss is set to $\left( \left\lbrack \mathbf{V}_{p} \right\rbrack_{bd} - \left\lbrack \mathbf{U}_{p}^{T}\ldots\mathbf{U}_{1}^{T}\mathbf{A} \right\rbrack_{bd} \right)^{2}$. $f(\mathbf{W},\gamma)$ is the soft weight operator.

\subsubsection{Pre-training}

We use a pre-training procedure consistent with that of DANMF and decompose the matrices $\mathbf{A}, \mathbf{V}_1, \mathbf{V}_2, \cdots, \mathbf{V}_{p-1}$ sequentially using the shallow NMF algorithm, which minimizes ${||\mathbf{A} - \mathbf{U}_1\mathbf{V}_1||}_{F}^2 + {||\mathbf{V}_1 - \mathbf{U}_1^T\mathbf{A}||}_{F}^2$ to obtain the pre-trained values of $\mathbf{U}_1 $ and $\mathbf{V}_1$, the same way $\mathbf{V}_1 \approx \mathbf{U}_2\mathbf{V}_2$, using
the NMF algorithm to minimize
${|| \mathbf{V}_1 - \mathbf{U}_{2}\mathbf{V}_2 ||}_F^2 + {||\mathbf{V}_2 - \mathbf{U}_2^T\mathbf{V}_1||}_{F}^2$
to get the pre-trained values of $\mathbf{U}_2$ and
$\mathbf{V}_2$, and so on, until all
$\mathbf{U}_i\text{\ and\ }\mathbf{V}_{i},\ \ i = 1,2,\cdots,p - 1$ factors are pre-trained.
\subsubsection{Optimization}

\textbf{AQ2}: we employ $f(\mathbf{W},\gamma)$ to answer \textbf{AQ2}. We update the weight $\mathbf{W}$ and matrix $\mathbf{U}_i$, $\mathbf{V}_i$ crosswise by optimizing (\ref{eq:11}) and (\ref{eq:12}):
\begin{equation}\label{eq:11}
\min_{1 \leq {\lbrack\mathbf{W}\rbrack}_{bd} \leq 0} \sum_{b,d}^{k,n}{\lbrack\mathbf{W}\rbrack}_{bd}l_{bd}   +f(\mathbf{W},\gamma)\\
\end{equation}
where 
\begin{equation}\nonumber
l_{bd} = \left( \left\lbrack \mathbf{V}_{p} \right\rbrack_{bd} - \left\lbrack \mathbf{U}_{p}^{T}\mathbf{U}_{p - 1}^{T}\ldots\mathbf{U}_{1}^{T}\mathbf{A} \right\rbrack_{bd} \right)^{2}
\end{equation}
\begin{equation}\label{eq:12}
\min_{\mathbf{U}_i,\mathbf{V}_p}\mathcal{L}=\mathcal{L}_D + \mathcal{L}_E^{'} + \lambda\mathcal{L}_{\text{reg}}
\end{equation}

\noindent \textbf{Update} $\mathbf{W}$ \textbf{:}

We employ alternative search strategy ~\cite{kumar:paced} to solve (\ref{eq:11}) and (\ref{eq:12}). In \textit{Silencer}, under fixed $\mathbf{U}$ and $\mathbf{V}$, $\mathbf{W}$ can be solved by (\ref{eq:13}). The update formula (\ref{eq:13}) for $\mathbf{W}$ are obtained from the soft weight operator (\ref{eq:9}) and the loss function (\ref{eq:11}).

\noindent\textbf{Update} $\mathbf{U}_i$ \textbf{and} $\mathbf{V}_p$ \textbf{:}



To obtain the update direction of $\mathbf{U}_{i}$, we fix all parameters other than $\mathbf{U}_{i}$ in (\ref{eq:12}) and use the Lagrange multiplier method to introduce the parameter $\mathbf{\Theta}$ to transform the formula into the unconditionally constrained (\ref{eq:17}):
\begin{equation}\label{eq:17}
\begin{gathered}
\min_{\mathbf{U}_{i},\mathbf{\Theta}_{i}}{\mathcal{L}(\mathbf{U}_{i},\mathbf{\Theta}_{i})}^{{}^{}}  = {\|\mathbf{A} - \mathbf{\Psi}_{i - 1}\mathbf{U}_{i}\mathbf{\Phi}_{i + 1}\mathbf{V}_{p}\|}_{F}^{2}  \\
+ \sum_{b,d = 1}^{k,n}{\left\{ \left [ \mathbf{W} \right ]  _{bd}\left( \left\lbrack \mathbf{V}_{p} \right\rbrack_{bd} - \left\lbrack \mathbf{\Phi}_{i + 1}^{T}\mathbf{U}_{i}^{T} \mathbf{\Phi}_{i - 1}^{T}\mathbf{A} \right\rbrack_{bd} \right)^{2}\right\}} \\
 + \text{tr}(\mathbf{\Theta}_{i}\mathbf{U}_{i}^{T}) +  f(\mathbf{W},\gamma) \\
\mathrm{\text{s.t.}}  \quad \mathbf{V}_{p} \geq \mathbf{0},\mathbf{U}_{i} \geq \mathbf{0}, \forall i = 1,2,\cdots,p,
\end{gathered}
\end{equation}
where $\mathbf{\Psi}_{i - 1} = \mathbf{U}_{1}\mathbf{U}_{2}\text{...}\mathbf{U}_{i - 1}$,$\ \mathbf{\Phi}_{i + 1} = \mathbf{U}_{i + 1}\mathbf{U}_{i}\text{...}\mathbf{U}_{p}$,
and $\mathbf{\Phi}_{p + 1}$, $\mathbf{\Psi}_{0}$ take the unit
matrix. By setting the partial derivative of
$\mathcal{L}(\mathbf{U}_{i}, \mathbf{\Theta}_{i})$ with
respect to $\mathbf{U}_i$ to \textbf{0}, we have (\ref{eq:18}):
\begin{equation}\label{eq:18}
\mathbf{\Theta}_{i} = - 2 \cdot \mathbf{\Psi}_{i - 1}^{T}\lbrack\mathbf{A}\mathbf{V}_{p}^{T} + \mathbf{A}{(\mathbf{W} \odot \mathbf{V_{p}})}^{T}\rbrack\mathbf{\Phi}_{i + 1}^{T} + 2 \cdot \mathbf{\Pi}_{i}
\end{equation} 
where
\begin{equation}\nonumber
\begin{aligned}
\mathbf{\Pi}_{i} &= \mathbf{\Psi}_{i - 1}^{T}\mathbf{\Psi}_{i - 1}\mathbf{U}_{i}\mathbf{\Phi}_{i + 1}\mathbf{V}_{p}\mathbf{V}_{p}^{T}\mathbf{\Phi}_{i + 1}^{T} \\
& + \mathbf{\Psi}_{i - 1}^{T}\mathbf{A}{(\mathbf{W} \odot \mathbf{\Phi}_{i + 1}^{T}\mathbf{U}_{i}^{T}\mathbf{\Psi}_{i - 1}^{T}\mathbf{A})}^{T}\mathbf{\Phi}_{i + 1}^{T}
\end{aligned}
\end{equation} 

Combining the complementary slackness condition $\mathbf{\Theta}_{i} \odot \mathbf{U}_{i} = \mathbf{0}$ from Karush-Kuhn-Tucker (KKT) conditions ~\cite{boyd2004convex}, we can infer to the update (\ref{eq:19}) for $\mathbf{U}_{i}$:
\begin{equation}\label{eq:19}
\mathbf{U}_{i} \longleftarrow \mathbf{U}_{i} \odot \frac{\mathbf{\Psi}_{i - 1}^{T}\lbrack\mathbf{A}\mathbf{V}_{p}^{T} + \mathbf{A}{(\mathbf{W} \odot \mathbf{\mathbf{V}_{p}})}^{T}\rbrack\mathbf{\Phi}_{i + 1}^{T}}{\mathbf{\Pi}_{i}}
\end{equation}
Similar to solving for $\mathbf{U}_{i}$ first fix all parameters other than $\mathbf{V}_{i}$ in (\ref{eq:12}) and obtain the updated (\ref{eq:20}) for $\mathbf{V}_{p}$ directly by deriving
\begin{equation}\label{eq:20}
\mathbf{V}_{p} \longleftarrow \mathbf{V}_{p} \odot \frac{\mathbf{\Psi}_{p}^{T}\mathbf{A} + \mathbf{W} \odot \mathbf{\Psi}_{p}^{T}\mathbf{A} + \lambda\mathbf{V}_{p}\mathbf{A}}{\mathbf{\Psi}_{p}^{T}\mathbf{\Psi}_{p}\mathbf{V}_{p} + \mathbf{W} \odot \mathbf{V}_{p} + \lambda\mathbf{V}_{p}\mathbf{D}}
\end{equation}

It is worth noting that observing the update formula reveals that $\mathbf{V}_{i}$ does not participate in the calculation except for $i=p$, which will be used so after pre-training
$\mathbf{V}_{i}$,$\ \ i = 1,2,\cdots,p - 1$ are not required to be updated again. The above has been reasoned to obtain all update formulas for optimizing \textit{Silencer}. We subsequently give the algorithmic procedure of \textit{Silencer} in Algorithm 1, where $m$ is the number of iterations.
\subsection{Complexity Analysis}

The algorithm is divided into two phases: the pre-training stage and fine-tuning stage. The computational complexity for the pre-training stage is of order $O(pt_p (n^2 r+nr^2))$, where $p$ is the number of layers, $t_p$ is the number of iterations to achieve convergence, and $r$ is the maximal layer size out of all layers. The computational complexity for the fine-tuning stage is of order $O(t_s pt_f (n^2 r+nr^2))$, where $O(pt_f (n^2 r+nr^2))$ is the time complexity of the fine-tuning phase of DANMF, $t_f$ and $t_s$ are the numbers of convergence of the DANMF algorithm and the number of self-paced learning convergence, respectively. In summary, the overall time complexity of this algorithm is $O(p(t_p+t_s t_f)(n^2 r+nr^2))$.

\subsection{Discussion}
By modifying the encoder of DANMF, we retain the excellent performance of the DANMF encoder in community detection and increase the robustness of community detection on noisy networks. We have tried to add a self-paced learning framework to handle the decoder, but the effect is poor. The main reason is that the network input by $\mathbf{V}_{p}$ of the encoder is relatively less noise and sparse. Thus, the weight added to the input data introduces more significant noise at first. The DANMF algorithm does not have complete convergence proof, and we will not prove it here.

\begin{table}[htb]
  \centering
  \caption{The layer sizes for six real-world networks with different noises.}
    \begin{tabular}{l|rrrcl}
    \toprule
    Dataset & \multicolumn{1}{l}{edges} & \multicolumn{1}{l}{nodes} & \multicolumn{1}{l}{Class} & \multicolumn{1}{l}{Noise} & Configuration \\
    \midrule
    Email & 25571 & 1005  & 42    & random & 1005-256-128-42 \\
    Cora  & 5429  & 2708  & 7     & random & 2708-256-64-7 \\
    Pubmed & 44338 & 19717 & 3     & random & 19717-512-64-3 \\
    Karate & 78    & 34    & 2     & Q-attack & 34-16-2 \\
    Football & 613   & 115   & 12    & Q-attack & 115-64-12 \\
    Polbooks & 441   & 105   & 3     & Q-attack & 105-64-32-3 \\
    \bottomrule
    \end{tabular}
  \label{tab:1}
\end{table}

\section{Experiments}
\subsection{Experimental Setup}
\subsubsection{Datasets and Evaluation Measures}

\begin{table*}[htbp]
  \centering
  \setlength\tabcolsep{3pt}
  \caption{Performance evaluation on Email, Cora, and Pubmed networks with different noise intensities.}
  \footnotesize
    \begin{tabular}{c|cc|ccc|cc|ccc|c}
    \toprule
    Metrics & Dataset & $p$     & NMF &DRANMF & DANMF & NE & ND & EdMot & RCD-GA & RCD-SE & \textit{Silencer} \\
    \midrule
    \multirow{8}[5]{*}{NMI} & \multirow{3}[2]{*}{Email} & 0     & 0.574±0.030 & 0.626±0.004 & 0.660±0.008 & 0.577±0.034 & 0.627±0.018 & 0.648±0.011 & 0.658±0.014 & 0.666±0.014 & \textbf{0.675±0.004} \\
          &       & 0.005 & 0.581±0.031 & 0.604±0.007 & 0.658±0.016 & 0.575±0.011 & 0.619±0.010 & 0.633±0.001 & 0.645±0.009 & 0.663±0.009 & \textbf{0.686±0.009} \\
          &       & 0.01  & 0.579±0.024 & 0.611±0.010 & 0.658±0.003 & 0.555±0.007 & 0.606±0.013 & 0.620±0.020 & 0.633±0.009 & 0.668±0.014 & \textbf{0.676±0.000} \\
\cmidrule{2-12}          & \multirow{3}[2]{*}{Cora} & 0     & 0.280±0.037 & 0.410±0.011 & 0.404±0.008 & 0.323±0.013 & 0.364±0.009 & 0.327±0.008 & 0.370±0.033 & 0.335±0.007 & \textbf{0.415±0.016} \\
          &       & 0.005 & 0.249±0.034 & 0.306±0.009 & 0.314±0.010 & 0.288±0.008 & 0.306±0.015 & 0.314±0.016 & 0.308±0.015 & 0.281±0.010 & \textbf{0.320±0.025} \\
          &       & 0.01  & 0.220±0.029 & 0.253±0.005 & 0.234±0.005 & 0.162±0.006 & 0.211±0.042 & 0.263±0.008 & 0.255±0.011 & 0.250±0.006 & \textbf{0.265±0.007} \\
\cmidrule{2-12}          & \multirow{2}[1]{*}{Pubmed} & 0     & 0.146±0.003 & 0.167±0.001 & 0.169±0.008 & 0.126±0.001 & 0.156±0.019 & 0.158±0.001 & -   & 0.138±0.004
 & \textbf{0.178±0.012} \\
          &       & 0.005 & 0.101±0.004 & 0.123±0.001 & 0.121±0.001 & 0.031±0.006 & 0.120±0.004 & 0.108±0.001 & -    & 0.104±0.016 & \textbf{0.129±0.001} \\
\midrule
    \multirow{8}[5]{*}{ARI} & \multirow{3}[1]{*}{Email} & 0     & 0.280±0.061 & 0.385±0.035 & 0.415±0.029 & 0.328±0.070 & 0.386±0.046 & 0.397±0.029 & 0.447±0.020 & 0.418±0.062 & \textbf{0.458±0.047} \\
          &       & 0.005 & 0.285±0.059 & 0.353±0.002 & 0.402±0.029 & 0.332±0.010 & 0.380±0.025 & 0.375±0.023 & 0.416±0.053 & 0.416±0.023 & \textbf{0.453±0.031} \\
          &       & 0.01  & 0.292±0.051 & 0.311±0.029 & 0.415±0.029 & 0.334±0.026 & 0.341±0.043 & 0.373±0.064 & 0.405±0.029 & 0.419±0.032 & \textbf{0.438±0.010} \\
\cmidrule{2-12}          & \multirow{3}[2]{*}{Cora} & 0     & 0.172±0.021 & 0.330±0.047 & 0.323±0.051 & 0.191±0.025 & 0.262±0.041 & 0.231±0.017 & 0.291±0.044 & 0.135±0.053 & \textbf{0.340±0.053} \\
          &       & 0.005 & 0.149±0.016 & 0.215±0.017 & 0.266±0.033 & 0.147±0.008 & 0.224±0.039 & 0.228±0.022 & 0.231±0.030 & 0.174±0.028 & \textbf{0.270±0.044} \\
          &       & 0.01  & 0.124±0.008 & 0.196±0.031 & 0.200±0.014 & 0.068±0.032 & 0.135±0.029 & 0.195±0.010 & \textbf{0.218±0.010} & 0.116±0.010 & 0.208±0.005 \\
\cmidrule{2-12}          & \multirow{2}[2]{*}{Pubmed} & 0     & 0.097±0.001 & 0.142±0.038 & 0.143±0.047 & 0.130±0.019 & 0.110±0.021 & 0.097±0.003 & -    & 0.089±0.006
& \textbf{0.167±0.072} \\
          &       & 0.005 & 0.090±0.001 & 0.120±0.009 & 0.124±0.015 & 0.020±0.021 & \textbf{0.139±0.021} & 0.108±0.023 & -   &  0.075±0.021 & 0.129±0.005 \\
    \midrule
    \multirow{8}[6]{*}{F1-score} & \multirow{3}[2]{*}{Email} & 0     & 0.310±0.055 & 0.410±0.034 & 0.437±0.028 & 0.365±0.061 & 0.413±0.044 & 0.420±0.027 & 0.469±0.020 & 0.449±0.057 & \textbf{0.479±0.035} \\
          &       & 0.005 & 0.315±0.055 & 0.380±0.003 & 0.425±0.029 & 0.368±0.007 & 0.408±0.025 & 0.398±0.020 & 0.440±0.050 & 0.447±0.022 & \textbf{0.474±0.030} \\
          &       & 0.01  & 0.320±0.047 & 0.341±0.028 & 0.436±0.029 & 0.367±0.024 & 0.371±0.038 & 0.398±0.060 & 0.428±0.027 & 0.450±0.030 & \textbf{0.460±0.010} \\
\cmidrule{2-12}          & \multirow{3}[2]{*}{Cora} & 0     & 0.329±0.024 & 0.453±0.038 & 0.447±0.038 & 0.363±0.016 & 0.396±0.032 & 0.363±0.009 & 0.411±0.038 & 0.202±0.083 & \textbf{0.463±0.046} \\
          &       & 0.005 & 0.312±0.020 & 0.353±0.018 & 0.395±0.031 & 0.344±0.012 & 0.363±0.033 & 0.376±0.029 & 0.363±0.024 & 0.300±0.021 & \textbf{0.398±0.042} \\
          &       & 0.01  & 0.293±0.015 & 0.334±0.027 & 0.342±0.014 & 0.283±0.009 & 0.286±0.026 & 0.338±0.009 & 0.346±0.005 & 0.209±0.019 & \textbf{0.364±0.016} \\
\cmidrule{2-12}          & \multirow{2}[2]{*}{Pubmed} & 0     & 0.461±0.006 & 0.491±0.020 & 0.493±0.025 & 0.444±0.024 & 0.495±0.017 & 0.482±0.014 & -   & 0.409±0.027 & \textbf{0.504±0.029} \\
          &       & 0.005 & 0.444±0.007 & \textbf{0.482±0.022} & 0.471±0.029 & 0.442±0.017 & 0.487±0.001 & 0.461±0.006 & -  & 0.364±0.027  & 0.478±0.035 \\
    \bottomrule
    \end{tabular}
  \label{tab:2}
\end{table*}

We use six real-world networks and two different noise treatments in our experiments. We provide an overview in Table \ref{tab:1}, where the three networks in the upper part are each applied with random noise to obtain three plots of noise probability $pn \in \{0,0.005,0.01\}$ to test the effectiveness of the algorithm under noisy inputs. In addition, we applied a modularity-based adversarial attack algorithm to the latter three networks, thus demonstrating that the method remains competitive even against data processed by the adversarial attack method, further illustrating the robustness of the \textit{Silencer}. More details about the network processing method will be shown in Section \ref{np}.

The six real networks selected in this paper all have the labels of real communities. Thus, we use three mainstream supervised metrics ~\cite{chakraborty2017metrics}: normalized mutual information (NMI), adjusted Rand index (ARI), and F1-score to evaluate the performance of community partitioning. In terms of the evaluation results, the larger the value of these three indicators, the better the partition effect. More details about metrics are shown in Appendix.

\begin{table}[tbp]
  \centering
  \caption{Decoder and Encoder errors are tested on the Email network. We calculated the decoder error $\frac{1}{n}\|\mathbf{A}-\mathbf{UV}\|_F$ and the decoder error $\frac{1}{n}\|\mathbf{V}-\mathbf{U}^{T}\mathbf{V}\|_F$ for the three algorithms \textit{Silencer}, DANMF, and DNMF, respectively, using the ten sets of $\mathbf{U}$ and $\mathbf{V}$ obtained by running in the aforementioned comprehensive experiments.}
    \begin{tabular}{l|rrr}
    \toprule
    $\mathcal{L}_d$    & 0     & 0.005 & 0.01 \\
    \midrule
    \textit{Silencer} & 0.137643 & 0.130926 & 0.130843 \\
    DANMF & 0.138539 & 0.138424 & 0.140126 \\
    DNMF  & 0.138246 & 0.123964 & 0.124229 \\
    \midrule
    $\mathcal{L}_E$    & 0     & 0.005 & 0.01 \\
    \midrule
    \textit{Silencer} & 0.047919 & 0.113661 & 0.140434 \\
    DANMF & 0.044149 & 0.044252 & 0.044854 \\
    DNMF  & 0.760229 & 4.332776 & 3.191598 \\
    \bottomrule
    \end{tabular}%
  \label{tab:3}%
\end{table}%
\subsubsection{Network processing method}
\label{np}
In real life, getting network data can be inadvertently or maliciously caused by noise. Here, we employ two different data processing methods corresponding to the two types of noise generation. The first one is to generate the added and removed edges with probability. The second one is to attack networks to produce noisy edges, which is used in \cite{chen2019ga,RobustECD}.

\begin{table*}[htbp]
  \centering
  \caption{The results of eight methods on three adversarial networks attacked by Q-attack.}
    \begin{tabular}{l|ccc|ccc|ccc}
    \toprule
    \multicolumn{1}{c|}{\multirow{2}[4]{*}{Methods}} & \multicolumn{3}{c|}{Karate} & \multicolumn{3}{c|}{Football} & \multicolumn{3}{c}{Polbooks} \\
\cmidrule{2-10}          & NMI   & ARI   & F1-score & NMI   & ARI   & F1-score & NMI   & ARI   & F1-score \\
    \midrule
    NMF   & 0.143±0.097 & 0.164±0.132 & 0.584±0.049 & 0.896±0.010 & 0.835±0.022 & 0.848±0.021 & 0.475±0.010 & 0.523±0.013 & 0.703±0.009 \\
    DNMF  & 0.250±0.087 & 0.113±0.105 & 0.619±0.022 & 0.902±0.005 & 0.848±0.024 & 0.860±0.022 & 0.522±0.011 & 0.543±0.027 & 0.715±0.011 \\
    NE & 0.141±0.031 & 0.116±0.021 & 0.576±0.008 & 0.804±0.027 & 0.653±0.049 & 0.683±0.044 & 0.543±0.037 & 0.578±0.050 & 0.755±0.034 \\
    ND & 0.208±0.122 & 0.177±0.161 & 0.600±0.079 & 0.886±0.028 & 0.822±0.056 & 0.837±0.051 & 0.517±0.014 & 0.553±0.038 & 0.730±0.022 \\
    DANMF & 0.335±0.054 & 0.209±0.073 & 0.636±0.020 & 0.903±0.016 & 0.848±0.036 & 0.861±0.034 & 0.539±0.020 & 0.542±0.016 & 0.716±0.006 \\
    EdMot & 0.185±0.043 & 0.220±0.054 & 0.602±0.025 & 0.891±0.022 & 0.818±0.064 & 0.833±0.058 & 0.517±0.020 & 0.569±0.039 & 0.730±0.027 \\
    RCD-GA & 0.249±0.027 & \textbf{0.303±0.032} & 0.642±0.016 & 0.899±0.006 & 0.841±0.007 & 0.854±0.006 & 0.496±0.014 & 0.584±0.019 & 0.739±0.013 \\
    RCD-SE &	0.226±0.022	& 0.260±0.039 &	0.628±0.013	& 0.894±0.010 & 0.816±0.015	& 0.831±0.014 & 0.560±0.025	& 0.614±0.049 &	0.778±0.038 \\
    \midrule
    \textit{Silencer} & \textbf{0.360±0.019} & 0.242±0.028 & \textbf{0.643±0.009} & \textbf{0.913±0.008} & \textbf{0.860±0.026} & \textbf{0.872±0.024} & \textbf{0.592±0.046} & \textbf{0.649±0.049} & \textbf{0.791±0.034} \\
    \bottomrule
    \end{tabular}
  \label{tab:4}
  \vskip -0.2in
\end{table*}

\begin{itemize}
\item
  \textbf{Random Noise}: we assume that each edge has the same probability of introducing noise. Here, the probability $p \in \{0,0.005,0.01\}$. Then every possible edge in each data set (whether it exists now or not) is assigned a random value $q$, $q \in [0,1]$.  we compare $q$ and $p$ for each edge, and the state of this edge is changed if $q \le p$ and unchanged if $q > p$. For example, there is a link between node $i$ and node $j$ in the original network, and now a noise of strength 0.01 is introduced to the original network, i.e., take $p = 0.01$. Suppose the edge between node $i$ and node $j$ is assigned $q = 0.005$. Then, since $q < p$, the state of the edge should be changed, i.e., there will be no link between nodes $i$ and $j$ after the noise is introduced.
\item
  \textbf{Q-attack}: It employs genetic algorithms to optimize the modularity-based objective function and then rewires the network to achieve state-of-the-art attacks. The attack number is set to 5\% of the total number of links in the network, and other parameters are not adjusted by referring to \cite{chen2019ga}.
\end{itemize}

\subsubsection{Baseline Methods}

The methods for robust community detection with link prediction are not employed as the baselines since their goal is the edge-missing case and not contains added edges.

\begin{itemize}
\item \textbf{NMF-based Community Detection}.
  \textbf{1) NMF}: By decomposing the feature matrix into the product of two low-dimensional matrix mapping matrices and the community membership matrix using a non-negative matrix, and thus finding the community where each node is located.\\
  \textbf{2) DANMF}: DANMF \cite{ye:autoencoder} is based on the NMF method and uses the idea of deep autoencoder to improve NMF for community discovery tasks.\\
  \textbf{3) DRANMF}: Deep robust auto-encoder nonnegative matrix factorization (DRANMF) \cite{esraa2023community} is an extension of DANMF against noise.\\
  \textbf{4) DNMF}: DANMF without an encoder.
\item
\textbf{Task-unaware Network Enhancement}.
  \textbf{1) ND}: The most related work in task-unaware network denoising for noisy networks via network deconvolution \cite{feizi2013network}. We conduct DANMF on the denoised network by ND.\\
  \textbf{2) NE}: A task-unaware network enhancement for weighted  networks \cite{wang2018network}. We conduct DANMF on the denoised network by NE.
\item
\textbf{Network Enhancement for Community Detection}.
  \textbf{1) EdMot}: EdMot \cite{EdMot2019} constructs the motif network through the adjacency matrix, then use the motif network to get the community and further get the possible edges. EdMot enhances the original adjacency matrix and obtains a better community discovery result.\\
  \textbf{2) RCD-GA}: Based on the modularity-based objective function, RCD-GA \cite{RobustECD} uses a specially designed evolutionary algorithm to obtain the enhanced adjacency matrix by adding or removing edges. RCD-GA then uses it to obtain better community discovery results.\\
  \textbf{3) RCD-SE}: The second framework presented in \cite{RobustECD}. RCD-SE is the network rewiring method.
\item
\textbf{\textit{Silencer}}: Unless otherwise specified, Silencer refers to DANMF as the optimizer.
\end{itemize}

\subsubsection{Parameters}

The EdMot, RCD-GA, RCD-SE, NE, and ND frameworks also employ DANMF as the underlying community discovery algorithm to ensure the fairness of the experiments. Throughout the experiments, the initialization settings, layer size configuration (Table \ref{tab:1}), update of DNMF, DRANMF, DANMF, and \textit{Silencer}\footnote{Code: https://github.com/SparseL/Silencer} times, and pre-training are kept consistent.
The regularization parameter $\lambda$ is adjustable in the range $\{10^{-3}, 10^{-2}, 10^{-1},10^{0}\}$, and the step size parameter $\eta$ is adjusted in $(1, 2.05]$. In addition, EdMot uses default parameters, and RCD-GA also uses default values for other parameters except for the two key parameters, the budget of edge addition $\beta _{a}$ and deletion $\beta _{d}$, which are controlled within $[0.01, 1]$. In
RCD-SE, we vary $\beta _{a}$
in $[0.01, 1]$. ND and NE use the default parameters. All algorithms are run ten times to obtain an average value representing their performance. We perform the Wilcoxon rank-sum test with $p-value \geq 0.05$ to validate the performance of \textit{Silencer}.

\subsection{Results}

\subsubsection{Case 1: Random Noise}

We obtain the results for three networks with different random noise intensity treatments, shown in Table \ref{tab:2}. Bold represents the best results among the eight community detection algorithms. Note that we do not show the results for the Pubmed network with $p=0. 01$ because all NMF-based test methods work poorly. Since RCD-GA runs for more than an hour on a noisy Pubmed network, we do not show its results.

Regarding the NMI metric, \textit{Silencer} outperforms the other algorithms in all cases. The ARI of \textit{Silencer} exceeds that of NMF, DRANMF, DANMF, EdMot, and RCD-SE in all cases, but \textit{Silencer} loses once to ND on the Pubmed network with $p=0.005$ and once to RCD-GA on the Cora network with $p=0.01$. In terms of the F1-score, \textit{Silencer} loses once to DRANMF on the Pubmed network, with a p-value of 0.005. Note that \textit{Silencer} is based on DANMF to silence the noise. \textit{Silencer} is better than DANMF on this setting. \textit{Silencer} outperforms RCD-SE in all cases. However, we found a fascinating phenomenon: \textit{in most cases, the performance of ND, NE, EdMot, RCD-SE, and RCD-GA is worse than that of DANMF, which indicates that these methods fail and reduce the performance of DANMF. RCD-SE has a positive gain in small networks but performs poorly in large networks. However, in all cases, our approach can improve DANMF performance.}

The combined comparison of the three metrics shows that \textit{Silencer} achieves optimal or suboptimal results in a quiet environment. In addition, \textit{Silencer} outperforms all other algorithms in noisy environments. The experimental results show that the rapid performance degradation of the DANMF model in noisy environments can be fully mitigated by embedding \textit{Silencer} framework in the encoder part, thus enhancing the robustness of the DANMF model in noisy environments.

A careful comparison of the effects of different intensities of noise on \textit{Silencer} reveals that, compared with DANMF, the greater the noise intensity, the stronger the noise resistance of \textit{Silencer}, just as the performance improvement ratio of NMI is higher with the increase of noise intensity. This phenomenon appears because the noise is randomly generated. The noise clips the inter-community edges or adds the intra-community edges.


We then analyze the reconstruction effect of the encoder and decoder and further explain why we should embed self-paced learning into the encoder. The tie values are shown in Table \ref{tab:3}. In the decoder part, the performance of the three algorithms is similar, among which \textit{Silencer} is slightly better, indicating that we have well inherited DANMF's ability to reconstruct the adjacency matrix. Our encoder's performance is moderate in the encoder part, but this justifies using the weight to avoid DANMF's over-reliance on the encoder by identifying noisy elements and applying small weights, thereby reducing layer-wise upscaling of the effect of post-learned noise.
\begin{figure}[ht]
 \centering
 \subfloat[]{\includegraphics[width=1.8in]{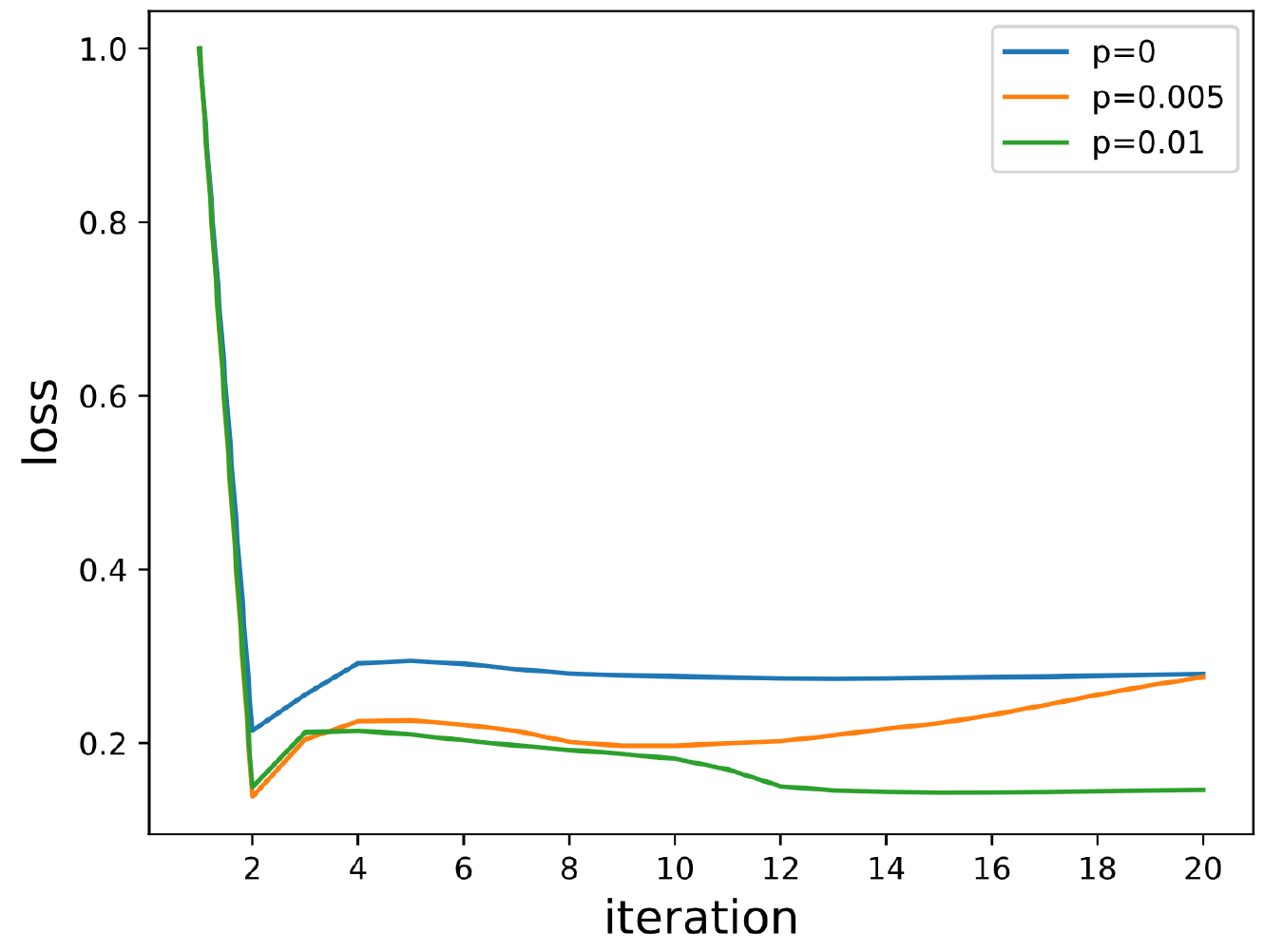}}
 \subfloat[]{\includegraphics[width=1.8in]{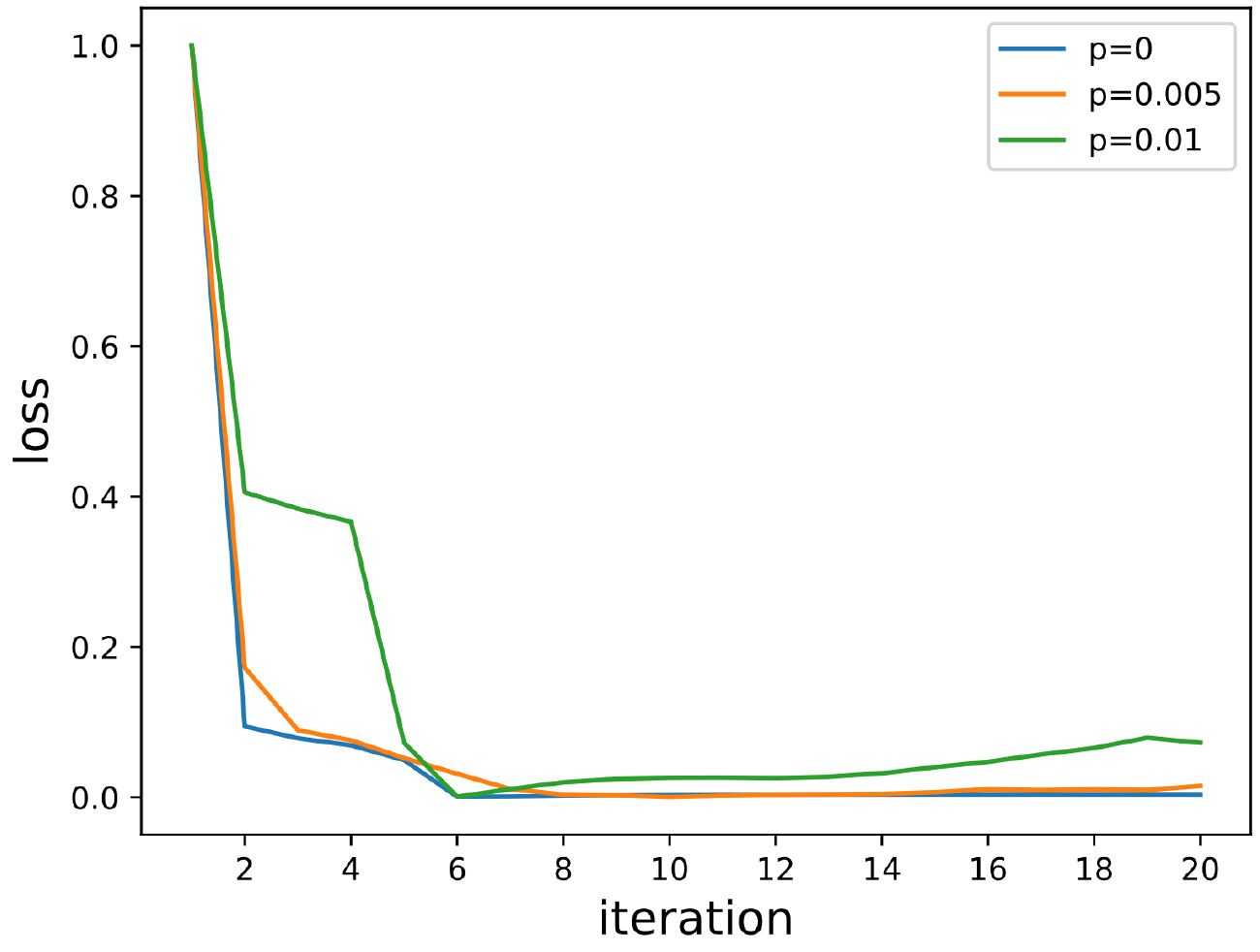}}\\
 \subfloat[]{\includegraphics[width=1.8in]{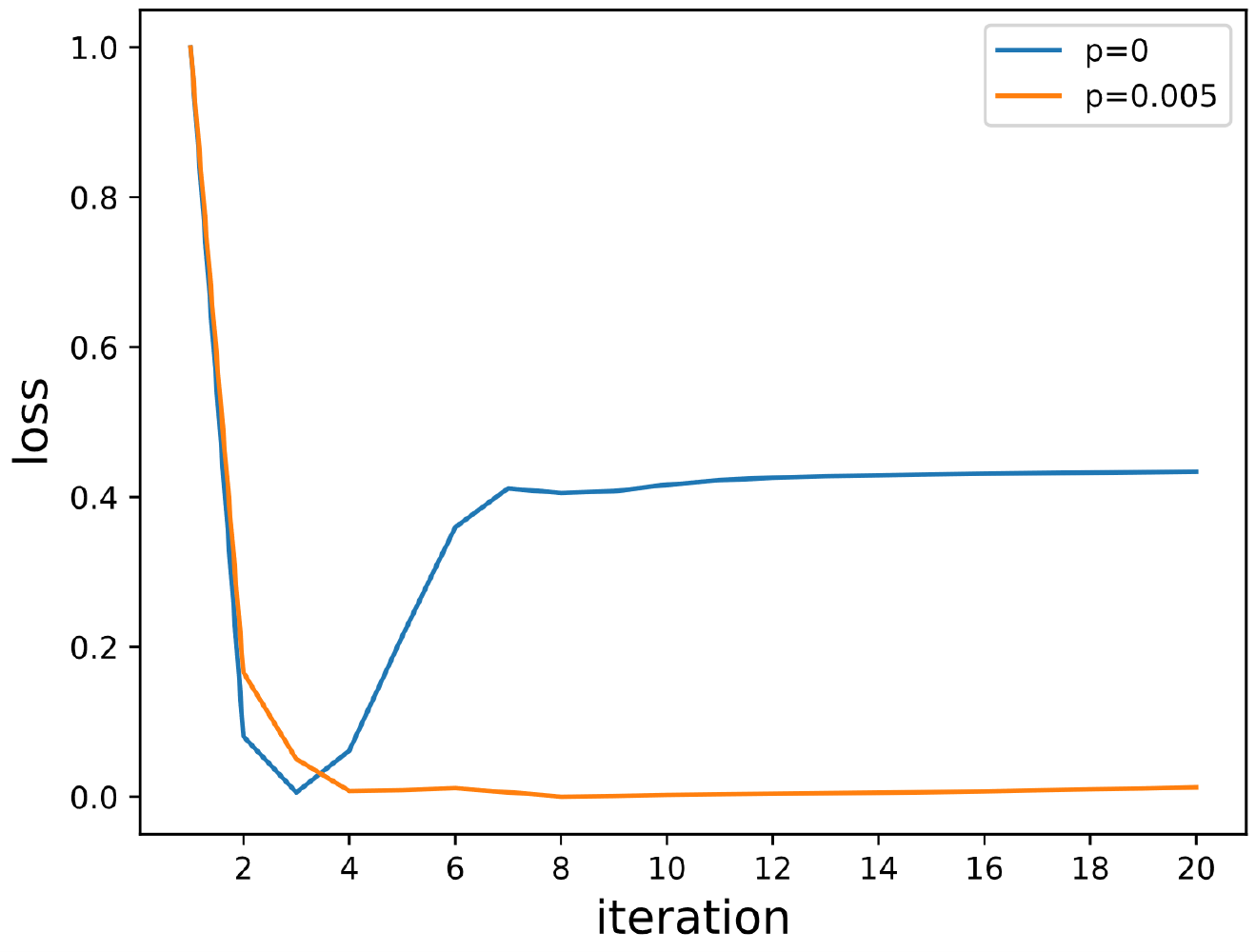}}
 \subfloat[]{\includegraphics[width=1.8in]{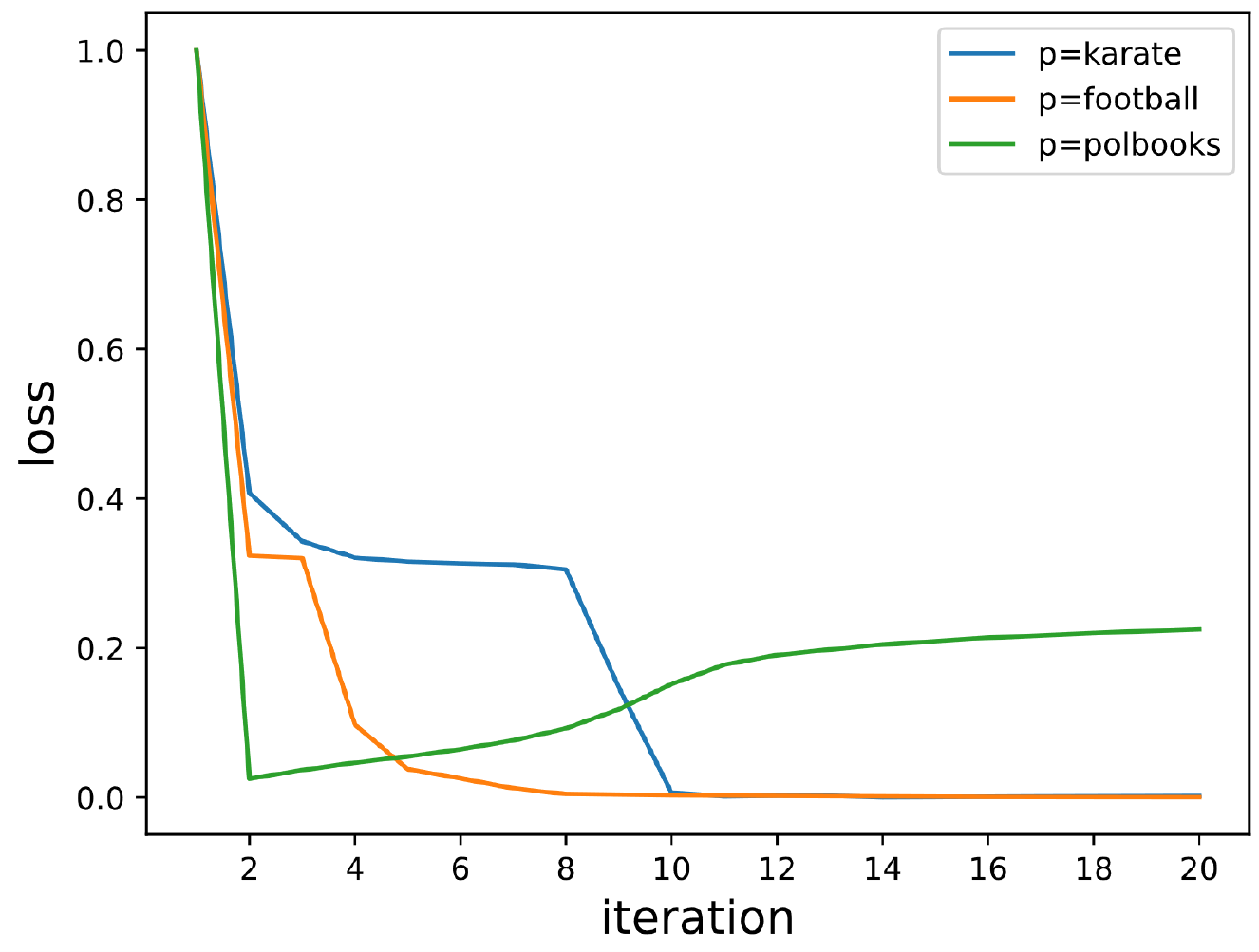}}\\
\caption{The reconstruction loss as a function of iterations for different networks. (a) Email network, (b) Cora network, (c) Pubmed network, and (d) Karate, Football, and Polbooks networks.}
\label{fig:3}
\end{figure}

\begin{figure}[htbp]
\centering
\includegraphics[width=3.4in]{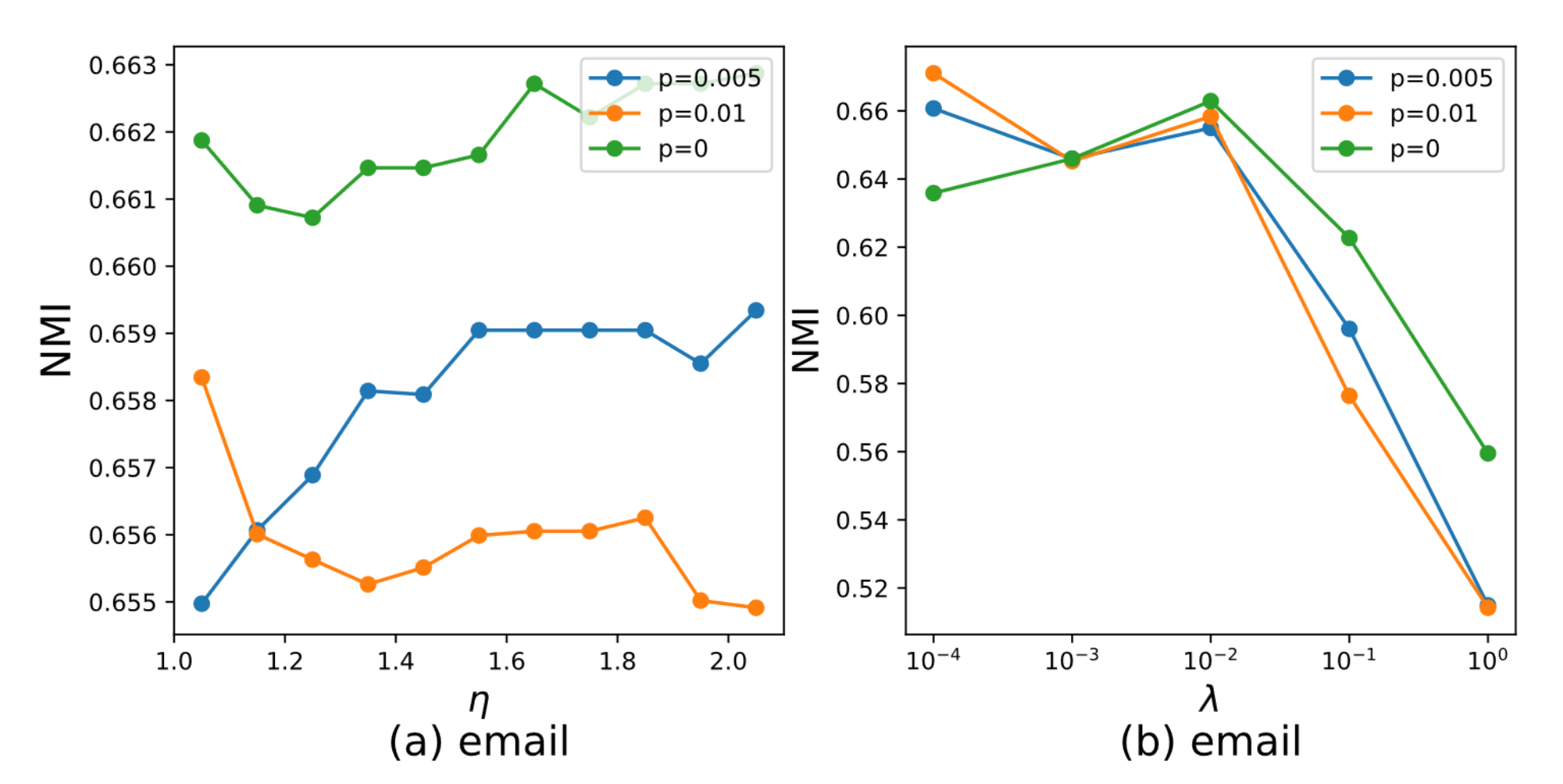}
\caption{Email network. (a) The effect of parameter $\eta$ on the performance of \textit{Silencer} under different noise intensities. (b) The effect of parameter $\lambda$ on the performance of \textit{Silencer} under different noise intensities.}
\label{fig:4}
\vskip -0.2in
\end{figure}

\begin{figure}[htbp]
\centering
\includegraphics[width=3.4in]{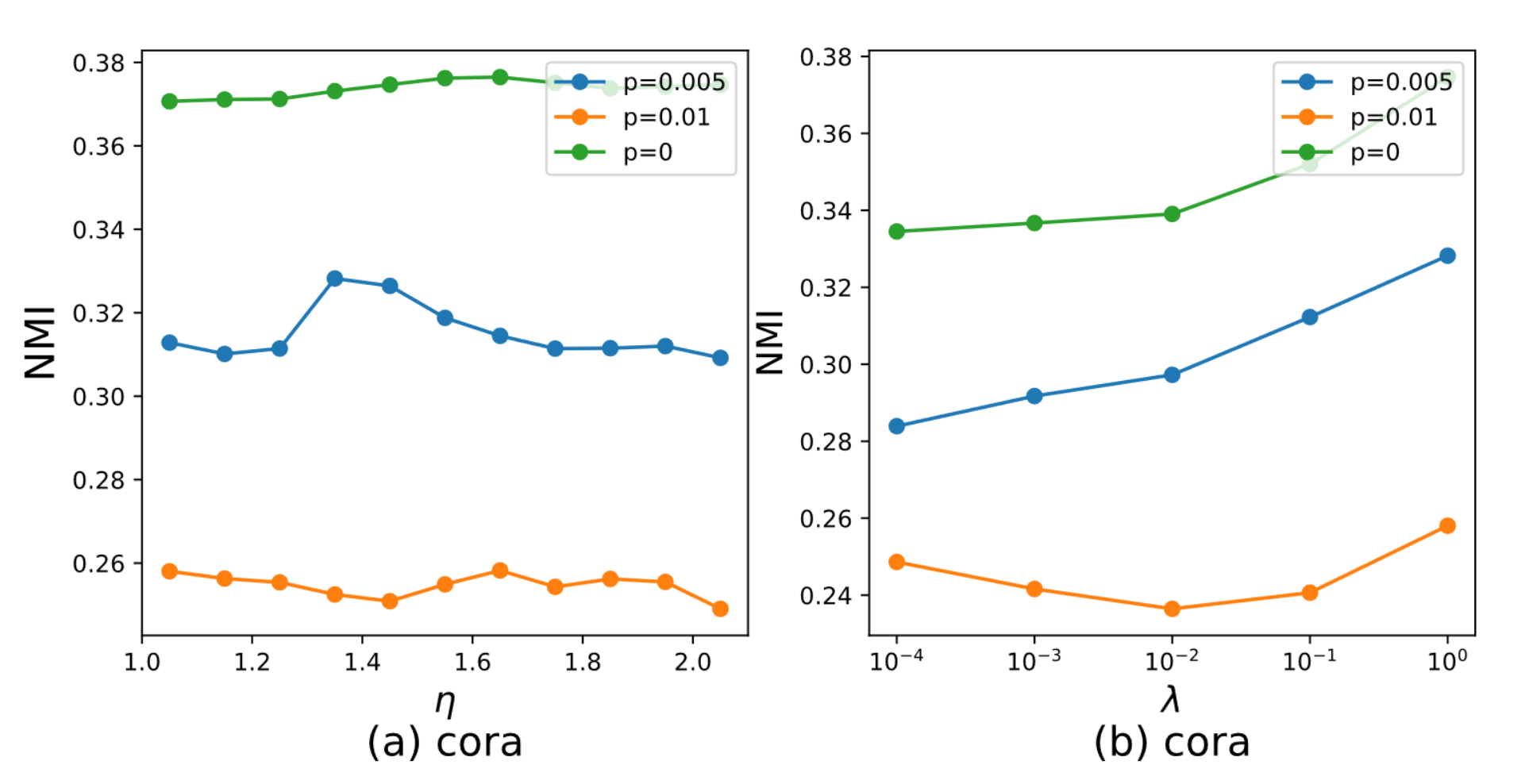}
\caption{Cora network. (a) The effect of parameter $\eta$ on the performance of \textit{Silencer} under different noise intensities. (b) The effect of parameter $\lambda$ on the performance of \textit{Silencer} under different noise intensities.}
\label{fig:5}
\end{figure}
\subsubsection{Case 2: Q-attack}

We obtain the results in Table \ref{tab:4} using the consistent experimental approach in Section 4.2.1 after using the Q-attack algorithm \cite{chen2019ga} for the three networks subjected to adversarial attacks. In terms of NMI and F1-score metrics, \textit{Silencer} outperforms the other algorithms in all cases. The ARI of \textit{Silencer} exceeds that of NMF, DNMF, DANMF, ND, NE, and EdMot in all cases, but \textit{Silencer} loses once to RCD-GA and RCD-SE on the Karate network.


We have compared network enhancement methods, such as NE and ND. The experimental results show that the gain of our scheme on DANMF is better than that of NE and ND on the Karate and Football networks. At the same time, NE and ND have a negative gain on DANMF. The performance of ND is better than DANMF on the Polbooks networks. NE outperforms DANMF in terms of the ARI and F1-score metrics but is worse than DANMF in terms of the NMI metric. The situation in EdMot is identical to that in NE. On the Karate and Polbooks networks, the NMI of RCD-GA is worse than that of DANMF, but RCD-GA outperforms DANMF in terms of ARI and F1-score metrics. Moreover, DANMF beats RCD-GA and RCD-SE on the Football network. The Football network has a more explicit community division than the Karate and Polbooks networks; the Football network has tighter intra-community and sparser inter-community links than the other networks.

The comparison of the three metrics shows that \textit{Silencer} still achieves the optimal results in the adversarial attack environment. Even though ARI and F1-score are not optimal in football, they are still suboptimal, and the difference is insignificant. At the same time, in the other two networks, our method improves very significantly, showing the strong robustness of \textit{Silencer}. The experimental results in Tables 1 and 2 show that the standard deviation of Silencer's results is relatively small. Therefore, different initializations will not significantly impact \textit{Silencer}, and \textit{Silencer} has strong stability.

\begin{table}[htbp]
  \centering
  \caption{The NMI improvement ratio of \textit{Silencer} with respect to DANMF on synthetic networks. The metric is calculated as follows: (Silencer's $Q$ value - DANMF's $Q$ value)/DANMF's $Q$ value.}
    \begin{tabular}{c|ccc}
    \toprule
    $p$     & ER    & WS    & BA \\
    \midrule
    0     & 42.49\% & 3.37\% & 9.67\% \\
    0.005 & 58.82\% & 0.15\% & 10.50\% \\
    0.01  & 35.29\% & 4.54\% & 16.46\% \\
    \bottomrule
    \end{tabular}
  \label{tab:5}
\end{table}

\subsubsection{Discussion}
In this section, we discuss the following question: \textit{what kinds of networks does Silencer work and fail?} We selected Erd{\H{o}}s–R{\'e}nyi random networks (ER) \cite{erdHos1960evolution}, Watts–Strogatz small-world networks (WS) \cite{watts1998collective}, and Barab{\'a}si–Albert scale-free networks (BA) \cite{barabasi1999emergence} representing most network characteristics for testing. Here, $N$=1000, using random noise ($p$={0, 0.005, 0.01}). The connection probability of the ER network is 0.1, the reconnection probability of the WS network is 0.5, and the number of edges added to the network each time in the BA network is 2. Since these networks have no ground truth community structure, modularity ($Q$) \cite{newman2004finding} is used as a performance metric.

The experimental results are shown in Table \ref{tab:5}. As the noise intensity increases, \textit{Silencer} still improves, demonstrating the ability of \textit{Silencer} to cope with solid noise. \textit{Silencer} has the most noticeable improvement on the ER network, followed by the BA network, and the worst on the WS network. However, our schemes all have performance improvements. In the ER network, the nodes are randomly arranged and connected, a network model with equal opportunities. The BA network is a network whose degree distribution of nodes conforms to a power-law distribution, and the degree of each node has a seriously uneven distribution. However, the WS network has the characteristics of "high network aggregation degree" and "low average path" simultaneously. In the order of ER, BA, and WS networks, the influence of changing a few connections on the community structure gradually becomes smaller, making it more challenging to identify the noisy edges. Therefore, the improvement of \textit{Silencer} is gradually insignificant.

\subsection{Convergence Analysis}

We display the reconstruction error with several iterations to illustrate the convergence rate of \textit{Silencer}. In Figure \ref{fig:3}, the loss function begins large and quickly shrinks with iterations until convergent for all networks after approximately 20 iterations ($m=20$). The loss function changes significantly during the descent, and the self-paced learning causes this fluctuation to produce new weights, as shown by a thorough comparison of each graph. \textit{Silencer} sacrifices some of the convergence rates compared to DANMF, which converges in about ten updates as stated in \cite{ye:autoencoder}, in order to increase robustness to noise.

\begin{figure}[t]
\centering
\includegraphics[width=3.4in]{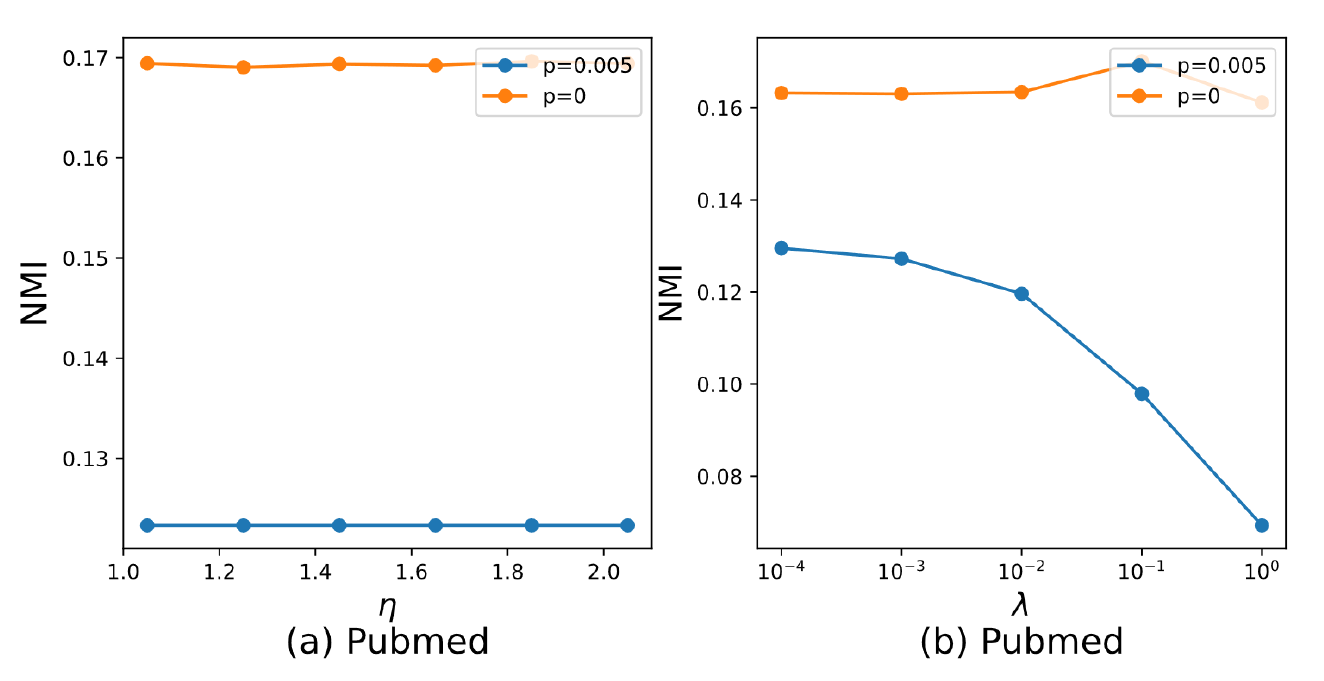}
\caption{Pubmed network. (a) The effect of parameter $\eta$ on the performance of \textit{Silencer} under different noise intensities. (b) The effect of parameter $\lambda$ on the performance of \textit{Silencer} under different noise intensities.}
\label{fig:6}
\vskip -0.2in
\end{figure}
\subsection{Parameter sensitivity}

We perform sensitivity tests on the parameter $\eta$ and the graph regular term adjustment parameter $\mathbf{\lambda}$ for \textit{Silencer}. $\lambda$ is tuned in the range of $\{10^{- 3},10^{- 2},10^{- 1},10^{0}\}$ and $\eta$ is tuned in the range of $(1, 2.05]$ in steps of 0.05. Figures \ref{fig:4}-\ref{fig:6} show the results for the Email, Cora, and Pubmed networks.

It can be seen that \textit{Silencer} is insensitive to changes in the $\eta$ parameter for different networks with various noise intensities.
In Figure \ref{fig:4}, the choice of parameter $\lambda$ has an essential impact on the performance of \textit{Silencer}. As the value of $\lambda$ increases, the NMI of \textit{Silencer} value first remains stable and then decreases. However, in Figure \ref{fig:5}, as the $\lambda$ value increases, the NMI of \textit{Silencer} rises. The difference in noise intensity does not affect the choice of optimal $\eta$ and $\lambda$ parameters since they have very similar trends. $\lambda = 0.01$ has relatively strong robustness. The network structure is also known to be an essential parameter. However, since the main purpose of this paper is to improve the robustness of DANMF for noisy networks, we do not conduct further experiments on this but directly follow the network structure in \cite{ye:autoencoder}.

\begin{figure}[htbp]
\centering
\includegraphics[width=3.5in]{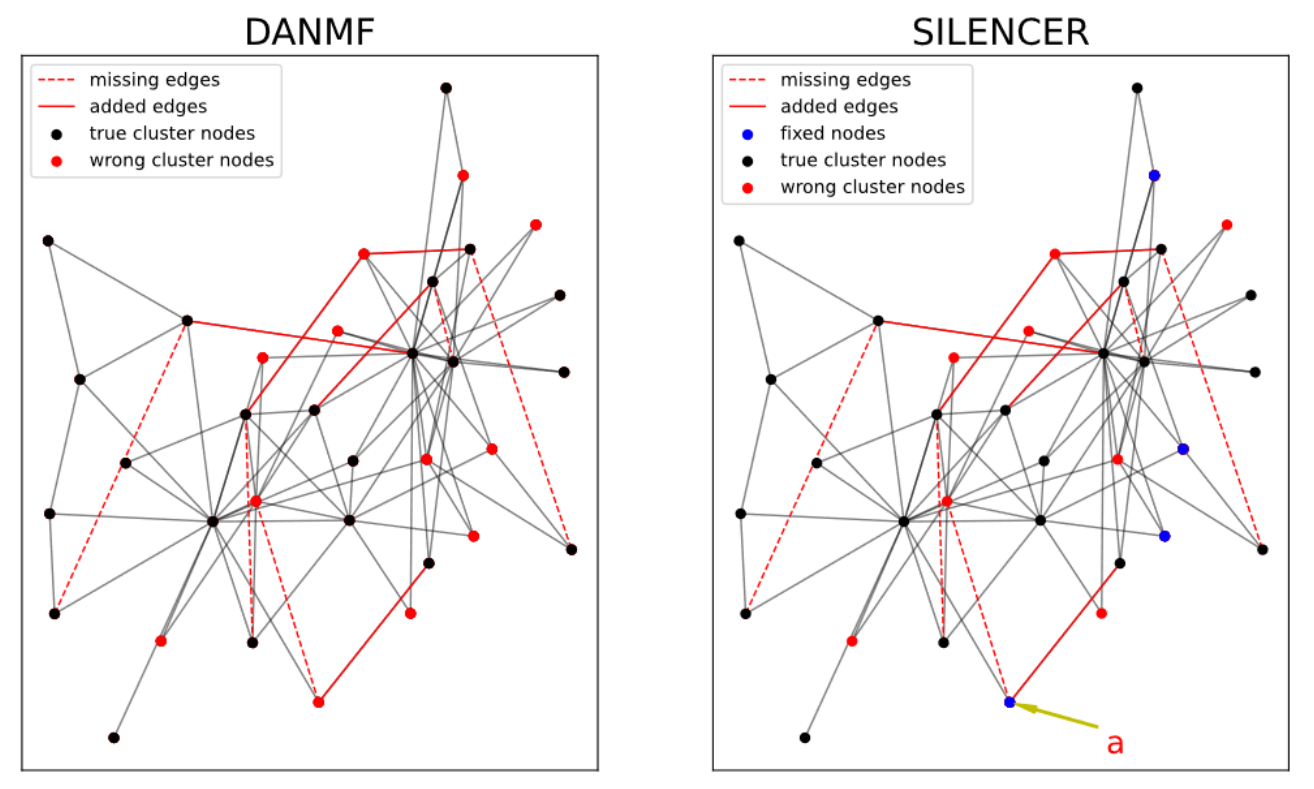}
\centering
\caption{Comparison of the results of the two algorithms on the noisy Karate network.}
\label{fig:7}
\vskip -0.1in
\end{figure}

\begin{figure}[htbp]
\centering
\includegraphics[width=3.5in]{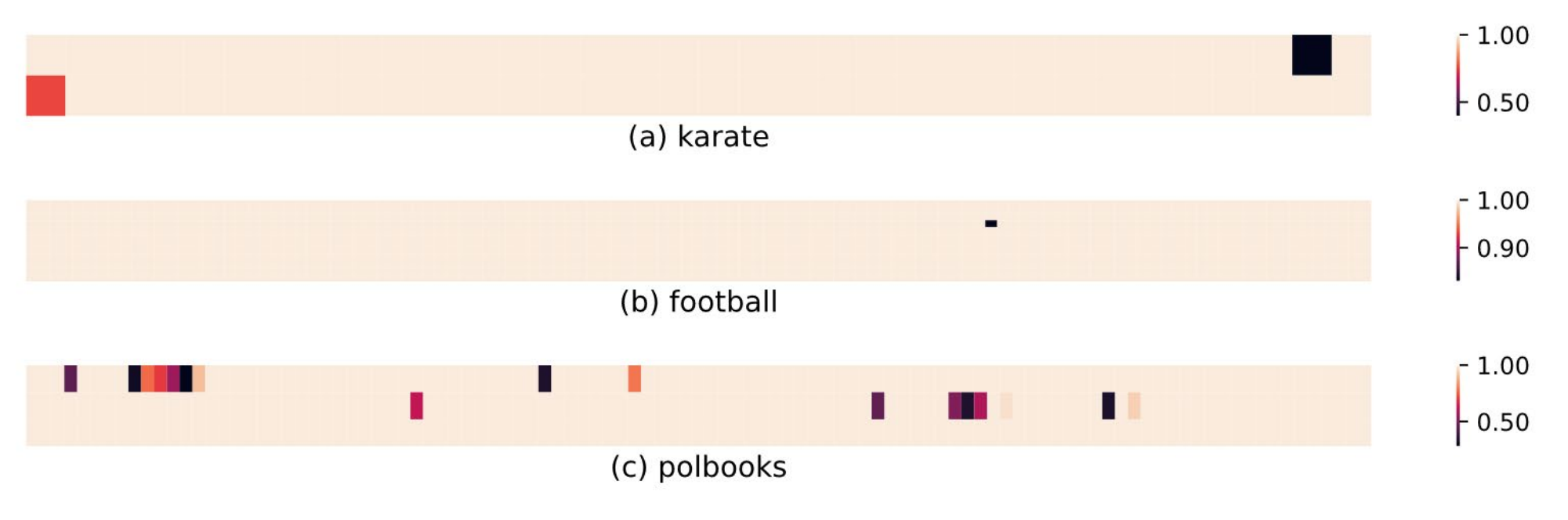}
\centering
\caption{The weight $\mathbf{W}$ in \textit{Silencer} on the three networks is represented as a heat map.}
\label{fig:8}
\vskip -0.1in
\end{figure}

\begin{table}[htbp]
  \centering
  \caption{The results of \textit{Silencer} and NMF for mixed noise. $\mathbf{A}$ is first processed using random noise.}
    \begin{tabular}{c|cc|l|l}
    \toprule
    Metrics & Dataset & $p$     & \multicolumn{1}{c|}{NMF} & \multicolumn{1}{c}{\textit{Silencer}} \\
    \midrule
    \multirow{8}[5]{*}{NMI} & \multirow{3}[2]{*}{Email} & 0     & 0.5457±0.0536 & \textbf{0.6157±0.0137} \\
          &       & 0.005 & 0.5585±0.0367 & \textbf{0.6194±0.0109} \\
          &       & 0.01  & 0.5447±0.0382 & \textbf{0.6096±0.0092} \\
\cmidrule{2-5}          & \multirow{3}[2]{*}{Cora} & 0     & 0.2764±0.0255 & \textbf{0.2952±0.0134} \\
          &       & 0.005 & 0.2323±0.0379 & \textbf{0.2691±0.0145} \\
          &       & 0.01  & 0.2010±0.0387 & \textbf{0.2297±0.0071} \\
\cmidrule{2-5}          & \multirow{2}[1]{*}{Pubmed} & 0     & 0.1418±0.0089 & \textbf{0.1495±0.0028} \\
          &       & 0.005 & 0.0975±0.0066 & \textbf{0.1048±0.0007} \\
 \midrule
    \multirow{8}[5]{*}{ARI} & \multirow{3}[1]{*}{Email} & 0     & 0.2559±0.0968 & \textbf{0.3936±0.0226} \\
          &       & 0.005 & 0.2627±0.0720 & \textbf{0.3883±0.0247} \\
          &       & 0.01  & 0.2450±0.0605 & \textbf{0.3663±0.0161} \\
\cmidrule{2-5}          & \multirow{3}[2]{*}{Cora} & 0     & 0.1931±0.0217 & \textbf{0.2023±0.0043} \\
          &       & 0.005 & 0.1653±0.0237 & \textbf{0.1722±0.0154} \\
          &       & 0.01  & \textbf{0.1342±0.0078} & 0.1281±0.0131 \\
\cmidrule{2-5}          & \multirow{2}[2]{*}{Pubmed} & 0     & 0.0861±0.0042 & \textbf{0.1339±0.0258} \\
          &       & 0.005 & 0.0766±0.0026 & \textbf{0.1129±0.0045} \\
    \midrule
    \multirow{8}[6]{*}{F1-score} & \multirow{3}[2]{*}{Email} & 0     & 0.2891±0.0909 & \textbf{0.4168±0.0220} \\
          &       & 0.005 & 0.2949±0.0669 & \textbf{0.4116±0.0243} \\
          &       & 0.01  & 0.2794±0.0562 & \textbf{0.3904±0.0154} \\
\cmidrule{2-5}          & \multirow{3}[2]{*}{Cora} & 0     & 0.3371±0.0103 & \textbf{0.3445±0.0124} \\
          &       & 0.005 & 0.3120±0.0139 & \textbf{0.3236±0.0200} \\
          &       & 0.01  & 0.2905±0.0134 & \textbf{0.2979±0.0048} \\
\cmidrule{2-5}          & \multirow{2}[2]{*}{Pubmed} & 0     & 0.4588±0.0097 & \textbf{0.4821±0.0176} \\
          &       & 0.005 & 0.4424±0.0106 & \textbf{0.4623±0.0134} \\
    \bottomrule
    \end{tabular}%
  \label{tab:nmf1}%
\end{table}%


\subsection{Visualization}

\begin{table*}[htbp]
  \centering
  \caption{The results of \textit{Silencer} and NMF for mixed noise. $\mathbf{A}$ is first processed using Q-attack.}
  \vskip -0.1in
    \begin{tabular}{c|ccc|ccc|ccc}
    \toprule
    \multirow{2}[4]{*}{Methods} & \multicolumn{3}{c|}{Karate} & \multicolumn{3}{c|}{Football} & \multicolumn{3}{c}{Polbooks} \\
\cmidrule{2-10}          & NMI   & ARI   & F1-score & NMI   & ARI   & F1-score & NMI   & ARI   & F1-score \\
\midrule
\cmidrule{2-10}    NMF   & 0.017±0.006 & -0.014±0.004 & 0.535±0.004 & 0.892±0.016 & 0.832±0.018 & 0.845±0.017 & 0.482±0.023 & 0.527±0.037 & 0.704±0.023 \\
    \textit{Silencer} & \textbf{0.034±0.007} & \textbf{-0.004±0.006} & \textbf{0.557±0.002} & \textbf{0.904±0.015} & \textbf{0.845±0.024} & \textbf{0.857±0.022} & \textbf{0.483±0.022} & \textbf{0.531±0.034} & \textbf{0.708±0.023} \\
    \bottomrule
    \end{tabular}
  \label{tab:nmf2}
  \vskip -0.2in
\end{table*}

Figure \ref{fig:7} shows the results of DANMF and \textit{Silencer} run on a noisy Karate network attacked by the Q-attack algorithm. The red edges indicate the attacked edges, where the dashed lines indicate the removed ones and the solid lines indicate the added ones. The nodes marked by red indicate the incorrectly clustered nodes, while the ones marked by blue in the middle panel are the nodes that are fixed by \textit{Silencer}. Node $a$ has two noisy edges, one added edge and one missing edge. DANMF has made the wrong community segmentation for node $a$, but Silencer can avoid the influence of noise edges and place node $a$ into the correct community. It is easy to see that these fixed nodes are directly or indirectly adjacent to the attacked edges. It can be intuitively seen that \textit{Silencer} is based on silencing the noisy edges to fix the wrongly clustered nodes.

In Figure \ref{fig:8}, we show the weight $\mathbf{W}$ of \textit{Silencer} when running on the three networks processed by the Q-attack algorithm to obtain the final clustering results. At the end of the iteration, we can see that there are still varying degrees of silence among the elements. Combined with the performance of \textit{Silencer} on various networks in Section V.B.2, we can intuitively see that the designed weight matrix directly affects the clustering ability of the algorithm. Especially in Figure \ref{fig:8}(b), the weight $\mathbf{W}$ is perturbed by only one element. However, combining with the results in Table \ref{tab:4}, we significantly improve the performance of DANMF. A small perturbation weight can significantly improve the performance of DANMF, which also confirms the feasibility of \textit{Silencer}. The positions of the lower weights are highly coincident with the true noisy edges.
\subsection{\textit{Silencer} in Mixed Noise}

Here, \textit{Silencer} refers to NMF as the optimizer. This section mainly tests the \textit{Silencer}'s ability to improve the network with strong mixed noise. In this more complicated scenario, network pollution is significantly more severe than that brought on by random noise and adversarial attacks. The definition of "Mixed Noise" is shown as follows:

\textbf{Definition 4 Mixed Noise:} \textit{$\mathbf{A}$ is first processed using one of the first two noise-adding methods. Then we use NMF to decompose $\mathbf{A}$ into $\mathbf{U}$ and $\mathbf{V}$. $\mathbf{UV}$ is the final noisy object.}

We establish two cases: 1) $\mathbf{A}$ is first processed using random noise; 2) $\mathbf{A}$ is first processed using Q-attack. The experimental results of \textit{Silencer} and NMF are shown in Tables \ref{tab:nmf1} and \ref{tab:nmf2}. The experimental results show that \textit{Silencer} outperforms NMF in all cases, demonstrating \textit{Silencer}'s strong denoising ability. For Case 2, \textit{Silencer}'s boosted ratio is less than in Case 1. Moreover, the mixed noise makes the polluted karate network very low in similarity with the original network, resulting in poor indicators. However, \textit{Silencer} still has performance improvements.

\section{Conclusion}
This paper proposes the \textit{Silencer} framework to address robust community detection from noisy networks via silencing noisy edges. We use NMF and DANMF methods to demonstrate the effectiveness of \textit{Silencer}. The experimental results have demonstrated the advantages of \textit{Silencer} in resisting three sizes of noise. We hope \textit{Silencer} will act as a solid baseline and help simplify future research on robust community detection. However, the existence of pixel-level loss leads to a reduction in the usage range of \textit{Silencer}. For example, modularity and motif-aware methods cannot be used because they cannot calculate the pixel-level loss. However, it can be well applied for deep community detection methods that can calculate pixel-level loss, such as \cite{chen2018supervised,Bo2020Structural}. Our work raises many open problems to further the development of robust community detection methods. For example, we only assume that the network edges are polluted and do not consider more complex node deletion or addition scenarios. We will develop an effective method to handle this case in future work.

\ifCLASSOPTIONcaptionsoff
  \newpage
\fi
\end{document}